\documentclass[aps,prb,twocolumn,showpacs,amsmath,amssymb,superscriptaddress,floatfix]{revtex4-2}
\usepackage[english]{babel}
\usepackage[utf8]{inputenc}
\usepackage{graphicx} 
\usepackage{bm} 
\usepackage{latexsym}
\usepackage{amsfonts}
\usepackage{bbold}
\usepackage{appendix}
\usepackage[colorlinks = true,
            linkcolor = blue,
            urlcolor  = blue,
            citecolor = blue,
            anchorcolor = blue]{hyperref}

\newcommand{\up}{\uparrow}
\newcommand{\down}{\downarrow}

\newcommand{\munu}{{\mu\nu}}

\newcommand{\bk}{\mathbf{k}}
\newcommand{\bq}{\mathbf{q}}
\newcommand{\bQ}{\mathbf{Q}}
\newcommand{\br}{\mathbf{r}}

\newcommand{\cA}{\mathcal{A}}
\newcommand{\cB}{\mathcal{B}}
\newcommand{\cG}{\mathcal{G}}
\newcommand{\cJ}{\mathcal{J}}
\newcommand{\cK}{\mathcal{K}}
\newcommand{\cP}{\mathcal{P}}
\newcommand{\cR}{\mathcal{R}}
\newcommand{\cS}{\mathcal{S}}

\newcommand{\dmu}{\partial_\mu}
\newcommand{\dnu}{\partial_\nu}
\newcommand{\Luv}{{\Lambda_\mathrm{uv}}}
\newcommand{\chit}{\widetilde{\chi}}
\renewcommand{\a}{a}
\renewcommand{\b}{b}

\newcommand{\tr}{{\rm tr}}
\newcommand{\Tr}{{\rm Tr}}

\graphicspath{{figures/}}
\begin{document}

\author{Pietro M. Bonetti}
\affiliation{Max Planck Institute for Solid State Research, Heisenbergstrasse 1, D-70569 Stuttgart, Germany}

\author{Walter Metzner}
\affiliation{Max Planck Institute for Solid State Research, Heisenbergstrasse 1, D-70569 Stuttgart, Germany}

\title{SU(2) gauge theory of the pseudogap phase in the two-dimensional Hubbard model}
\date{\today}
\begin{abstract}
We present a SU(2) gauge theory of fluctuating magnetic order in the two-dimensional Hubbard model. The theory is based on a fractionalization of electrons in fermionic chargons and bosonic spinons. The chargons undergo N\'eel or spiral magnetic order below a density dependent transition temperature $T^*$. Fluctuations of the spin orientation are described by a non-linear sigma model obtained from a gradient expansion of the spinon action. The spin stiffnesses are computed from a renormalization group improved random phase approximation. Our approximations are applicable for a weak or moderate Hubbard interaction. The spinon fluctuations prevent magnetic long-range order of the electrons at any finite temperature. The phase with magnetic chargon order exhibits many features characterizing the pseudogap regime in high-$T_c$ cuprates: a strong reduction of charge carrier density, a spin gap, Fermi arcs, and electronic nematicity.
\end{abstract}
\pacs{}
\maketitle
\section{Introduction}

Besides their exceptionally high transition temperatures for superconductivity, a peculiar and fairly universal feature of hole-doped cuprate superconductors is their pseudogap behavior for temperatures above $T_c$, observed in a broad doping range from the underdoped into the optimally doped regime \cite{Keimer2015, Proust2019}. The pseudogap behavior sets in at a temperature $T^*$ which is much higher than $T_c$ in the underdoped regime, and merges with $T_c$ near optimal doping. It is characterized by a spin gap, a reduction of the charge carrier concentration, a suppression of the electronic density of states, a gap for single-particle excitations in the antinodal region of the Brillouin zone, and a reconstructed Fermi surface which in photoemission looks like Fermi arcs. It is also associated with a tendency to electronic nematicy, where the electronic state breaks the tetragonal symmetry of the crystal. Suppressing superconductivity by high magnetic fields, the pseudogap regime extends into the region which in the absence of magnetic fields is superconducting -- for doping concentrations as high as about 20 percent \cite{Proust2019}. 

There is convincing numerical evidence for pseudogap behavior in the strongly interacting two-dimensional Hubbard model, in particular from quantum cluster calculations \cite{Qin2022}. An unbiased ``fluctuation diagnostics'' \cite{Gunnarsson2015} of the contributions to the self-energy has revealed that the pseudogap is generated predominantly by antiferromagnetic fluctuations, that is, by spin fluctuations with wave vectors at or near $(\pi,\pi)$. 

While the guidance provided by numerical results is clearly very valuable, a deeper understanding of the pseudogap phenomenon and data with a higher momentum resolution remain desirable. The momentum resolution of the self-energy and of other momentum dependent quantities is limited in all cluster methods, since the computational effort grows exponentially with the cluster size. Also long-range correlations (in real space) beyond the cluster size cannot be captured. The same limitations hold of course for direct numerical simulations of finite systems.

In this situation approximate analytic theories can provide further insights, especially concerning long-range correlations and the fine-structure in momentum space. Early theories of the pseudogap phenomenon were based on weak-coupling diagrammatic perturbation expansions, most notably Moriya's renormalized theory \cite{Moriya2000} and the two-particle self-consistent theory by Vilk and Tremblay \cite{Vilk1996}. The Mermin-Wagner theorem on the absence of spin symmetry breaking at finite temperatures is respected in these theories, but the pseudogap seems to develop only for fairly large magnetic correlation lengths, while the numerical data show that strong short-ranged correlations are sufficient.

More recently it was shown by Sachdev, Scheurer, and coworkers that many features of the pseudogap behavior observed in cuprates can be captured by a SU(2) gauge theory \cite{Sachdev2016, Chatterjee2017, Scheurer2018, Wu2018, Sachdev2019}. This approach is based on a fractionalization of the electron into a fermionic ``chargon'' and a charge neutral ``spinon''. The latter is a SU(2) matrix providing a space and time dependent local reference frame \cite{Schulz1995}. The local spin rotations can be parametrized by a SU(2) gauge field, and the fractionalization leads to a gauge redundancy. One can then consider states where the chargons exhibit some sort of magnetic order (for example, N\'eel or spiral), while the spinon fluctuations prevent symmetry breaking and magnetic long-range order of the physical spin-carrying electrons \cite{Dupuis2002, Borejsza2004, Sachdev2009}. Quantities involving only charge degrees of freedom behave essentially as in a conventional magnetically ordered state, and the Fermi surface gets correspondingly reconstructed. While long-range order is absent, at least at finite temperatures, the electrons are subject to a ``topological'' order in the sense that smoothly varying local spin rotations can map the fluctuating spin configurations to an ordered pattern, that is, there is no proliferation of topological defects \cite{Sachdev2019}. 

In this paper we formulate a SU(2) gauge theory for the fluctuating antiferromagnet in a way that allows us to compute, in a decent approximation for weak or moderate Hubbard interactions, effective low-energy parameters and physical quantities as a function of the microscopic model parameters. The chargon order parameter is computed from a renormalized mean-field theory \cite{Wang2014} which takes high energy (above $T^*$) spin, charge, and pairing fluctuations into account on equal footing. We allow for N\'eel and planar spiral order with generally incommensurate ordering wave vectors. The spinon dynamics is described by a non-linear sigma model (NL$\sigma$M). The parameters of the NL$\sigma$M, that is, the spin stiffnesses, are computed from a renormalized RPA for the SU(2) gauge field response of the chargons \cite{Bonetti2022a}, and the ultraviolet cutoff is estimated via the magnetic coherence length. The NL$\sigma$M is evaluated in a large $N$ expansion.
Applying the general theory to the Hubbard model with next and next-nearest neighbor hopping at a moderate interaction strength (about half band width), we obtain a broad finite temperature pseudogap regime on the hole doped side and a narrower pseudogap region for electron doping. Nematic order is present at sufficiently low temperatures for hole doping, but not for electron doping. There is no magnetic long-range order at $T>0$, in agreement with the Mermin-Wagner theorem, and the spin excitations are gapped.
The spinon quantum fluctuations are not strong enough to destroy magnetic long-range order in the ground state, except possibly near the edge of the pseudogap regime at large hole doping. In the hole doped pseudogap regime, the Fermi surfaces extracted from the single-particle spectral function have the form of hole pocket boundaries with a truncated back side. Their topology is thus the same as for the experimentally observed Fermi arcs.  

The paper is structured as follows. In Sec.~II we derive the general structure of the SU(2) gauge theory for the pseudogap phase with N\'eel or spiral order in the chargon sector. In Sec.~III we describe how we compute the parameters of the gauge theory, in particular the spin stiffnesses, from the underlying microscopic model. Sec.~IV deals with the solution of the nonlinear sigma model for the spinon fluctuations in a large $N$ expansion. Results for the two-dimensional Hubbard model are presented in Sec.~V. We conclude with a summary and a final discussion of our theory in Sec.~VI.


\section{SU(2) gauge theory} 
\label{sec: SU(2) gauge theory}

\subsection{Fractionalizing the electron field}
We consider the Hubbard model on a square lattice with units of length such that the lattice spacing is one. The Hubbard action in imaginary time reads
\begin{eqnarray}
 \mathcal{S}[c,c^*] &=& 
 \int_0^\beta\!d\tau \bigg\{ \sum_{j,j',\sigma} c^*_{j\sigma}
 \left[ \left( \partial_\tau - \mu\right)\delta_{jj'} + t_{jj'} \right] c_{j'\sigma}
 \nonumber \\
 && + \; U \sum_j n_{j\up}n_{j\down} \bigg\} , 
 \label{eq: Hubbard action}
\end{eqnarray}
where $c_{j\sigma} = c_{j\sigma}(\tau)$ and $c^*_{j\sigma} = c^*_{j\sigma}(\tau)$ are Grassmann fields corresponding to the annihilation and creation, respectively, of an electron with spin orientation $\sigma$ at site $j$, and $n_{j\sigma} = c^*_{j\sigma} c_{j\sigma}$. The chemical potential is denoted by $\mu$, and $U > 0$ is the strength of the (repulsive) Hubbard interaction. To simplify the notation, we write the dependence of the fields on the imaginary time $\tau$ only if needed for clarity.

The action in \eqref{eq: Hubbard action} is invariant under \emph{global} SU(2) rotations acting on the Grassmann fields as 
\begin{equation} \label{eq: SU(2) transf. electrons}
 c_j \to \mathcal{U} c_j, \quad
 c^*_j \to c^*_j \, \mathcal{U}^\dagger,
\end{equation}
where $c_j$ and $c^*_j$ are two-component spinors composed from $c_{j\sigma}$ and $c^*_{j\sigma}$, respectively, while $\mathcal{U}$ is a SU(2) matrix acting in spin space.

To separate collective spin fluctuations from the charge degrees of freedom, we fractionalize the electronic fields as~\cite{Schulz1995, Dupuis2002, Borejsza2004, Sachdev2009}
\begin{equation} \label{eq: electron fractionaliz.}
 c_j = R_j \, \psi_j , \quad
 c^*_j = \psi^*_j \, R^\dagger_j ,
\end{equation}
where $R_j \in \mbox{SU(2)}$, to which we refer as ``spinon'', is composed of bosonic fields, and the components $\psi_{js}$ of the ``chargon'' spinor $\psi_j$ are fermionic. According to \eqref{eq: SU(2) transf. electrons} and \eqref{eq: electron fractionaliz.} the spinons transform under the global SU(2) spin rotation by a \emph{left} matrix multiplication, while the chargons are left invariant. Conversely, a U(1) charge transformation acts only on $\psi_j$, leaving $R_j$ unaffected.
We have therefore separated the spin degrees of freedom of the physical electrons, now encoded in the spinons, from their charge, carried by the chargons.
The transformation in Eq.~\eqref{eq: electron fractionaliz.} introduces a redundant SU(2) gauge symmetry, acting as 
\begin{subequations} \label{eq: gauge symmetry}
\begin{align}
 & \psi_j \to \mathcal{V}_j\,\psi_j , \quad
   \psi^*_j \to \psi^*_j \, \mathcal{V}^\dagger_j , \\
 & R_j \to R_j \, \mathcal{V}_j^\dagger, \quad
   R^\dagger_j \to \mathcal{V}_j \, R^\dagger_{j} ,
\end{align}
\end{subequations}
with $\mathcal{V}_j\in \mbox{SU(2)}$.
Hence, the components $\psi_{js}$ of $\psi_j$ carry a SU(2) gauge index $s$, while the components $R_{j,\sigma s}$ of $R_j$ have two indices, the first one ($\sigma$) corresponding to the global SU(2) symmetry, and the second one ($s$) to SU(2) gauge transformations. 

We now rewrite the Hubbard action in terms of the spinon and chargon fields. The quadratic part of \eqref{eq: Hubbard action} can be expressed as \cite{Borejsza2004}
\begin{eqnarray}
 \mathcal{S}_0[\psi,\psi^*,R] &=& \int_0^\beta\!d\tau
 \bigg\{ \sum_j \psi^*_j \left[ \partial_\tau - \mu - A_{0,j} \right]
 \psi_{j} \nonumber \\
 && + \, \sum_{j,j'}t_{jj'}\,\psi^*_{j}\, e^{-\mathbf{r}_{jj'} \cdot \left(\boldsymbol{\nabla} - i\mathbf{A}_j \right)} \, \psi_j \bigg\},
\label{eq: S0 chargons spinons}
\end{eqnarray}
where we have introduced a SU(2) gauge field, defined as 
\begin{equation}
    A_{\mu,j} = (A_{0,j},\mathbf{A}_j) = i R^\dagger_j \dmu R_j,
    \label{eq: gauge field definition}
\end{equation}
with $\dmu = (i\partial_\tau,\boldsymbol{\nabla})$. Here, the nabla operator $\boldsymbol{\nabla}$ is defined as generator of translations on the lattice, that is,
$e^{-\br_{jj'}\cdot\boldsymbol{\nabla}}$ with $\br_{jj'} = \br_j - \br_{j'}$
is the translation operator from site $j$ to site $j'$.

To rewrite the interacting part in \eqref{eq: Hubbard action}, we use the decomposition~\cite{Weng1991,Schulz1995,Borejsza2004}
\begin{equation}
 n_{j\up}n_{j\down} = \frac{1}{4}(n_j)^2 - \frac{1}{4}(c^*_j \,
 \vec{\sigma}\cdot\hat{\Omega}_j \, c_j)^2,
\label{eq: interaction decomposition}
\end{equation}
where $n_j = n_{j,\up} + n_{j,\down}$ is the charge density operator,
$\vec{\sigma} = (\sigma^1,\sigma^2,\sigma^3)$ are the Pauli matrices, and $\hat{\Omega}_j$ is an arbitrary time- and site-dependent unit vector. Inserting the decomposition \eqref{eq: electron fractionaliz.}, the interaction term of the Hubbard action can therefore be written as
\begin{equation}
 \mathcal{S}_\mathrm{int}[\psi,\psi^*,R] =
 \int_0^\beta\!d\tau \, U \sum_j \left[\frac{1}{4} (n_j^\psi)^2 -
 \frac{1}{4}(\vec{S}^\psi_j\cdot\hat{\Omega}^R_j)^2 \right] ,
\end{equation}
where $n^\psi_j = \psi^*_j\psi_j$ is the chargon density operator, $\vec{S}^\psi_j = \frac{1}{2} \psi^*_j\vec{\sigma}\psi_j$ is the chargon spin operator, and $\hat\Omega^R_j$ is a unit vector obtained by rotating $\hat\Omega_j$ as
\begin{equation}
 \vec\sigma \cdot \hat\Omega^R_j =
 R^\dagger_j \, \vec\sigma \cdot \hat\Omega_j \, R_j .
\label{eq: Omega and Omega^R}
\end{equation}
Using \eqref{eq: interaction decomposition} again, we obtain 
\begin{equation} \label{eq: S_int final}
 \mathcal{S}_\mathrm{int}[\psi,\psi^*,R] =
 \int_0^\beta\!d\tau \, U \sum_j n^\psi_{j\up}n^\psi_{j\down},
\end{equation}
with $n^\psi_{js} = \psi^*_{js} \psi_{js}$.
Therefore, the final form of the action $\mathcal{S} = \mathcal{S}_0 + \mathcal{S}_\mathrm{int}$ is nothing but the Hubbard model action where the physical electrons have been replaced by chargons coupled to a SU(2) gauge field.

Since the chargons do not carry the physical spin degree of freedom, a global breaking of their SU(2) gauge symmetry ($\langle \vec{S}^\psi_j \rangle \neq 0$) does not necessarily imply long range order for the electrons.
The matrices $R_j$ describe directional fluctuations of the order parameter $\langle \vec{S}_j \rangle$, where, at low temperatures, the most important ones vary slowly in space and time. 


\subsection{Nonlinear sigma model}
\label{sec: NLsM}
We now derive a low energy effective action for the spinon fields $R_j$ by integrating out the chargons,
\begin{equation}
 e^{-\mathcal{S}_\mathrm{eff}[R]} =
 \int\! \mathcal{D} \psi \mathcal{D} \psi^* \,
  e^{-\mathcal{S}[\psi,\psi^*,R]} . 
\label{eq: integral over psi}
\end{equation}
Since the action $\mathcal{S}$ is quartic in the fermionic fields, the functional integral must be carried out by means of an approximate method. In previous works~\cite{Schulz1995,Dupuis2002,Borejsza2004} a Hubbard-Stratonovich transformation has been applied to decouple the chargon interaction, together with a saddle point approximation on the auxiliary bosonic (Higgs) field. We will employ an improved approximation based on the functional renormalization group \cite{Metzner2012}, which we describe in Sec.~\ref{sec: fRG+MF}. 

The effective action for the spinons can be obtained by computing the response functions of the chargons to a fictitious SU(2) gauge field. Since we assign only low energy long wave length fluctuations to the spinons in the decomposition \eqref{eq: electron fractionaliz.}, the spinon field $R_j$ is slowly varying in space and time. Hence, we can perform a gradient expansion. To second order in the gradient $\dmu R_j$, the effective action $\mathcal{S}_\mathrm{eff}[R]$ has the general form
\begin{equation}
 \mathcal{S}_\mathrm{eff}[R] = \int_\mathcal {T} \! dx \Big[
 \cB^a_\mu A_{\mu}^a(x) +
 \textstyle{\frac{1}{2}} \cJ^{ab}_\munu  
 A_{\mu}^a(x) A_{\nu}^b(x) \Big] ,
\label{eq: effective action Amu}
\end{equation}
where $x = (\tau,\br)$ combines imaginary time and space coordinates, $\mathcal{T} = [0,\beta] \times \mathbb{R}^2$ is the integration region, and repeated indices are summed. We have expanded the gauge field $A_\mu$ in terms of the SU(2) generators,
\begin{equation}
 A_\mu(x) = A_\mu^a(x) \, \sigma^a/2 ,
\label{eq: Amu SU(2) generators}
\end{equation}
with $a$ running from 1 to 3. 
In line with the gradient expansion, the gauge field is now defined over a \emph{continuous} space-time. The coefficients in~\eqref{eq: effective action Amu} do not depend on $x$ and are given by
\begin{eqnarray} \label{eq: def Ba}
 \cB^a_\mu &=& \frac{1}{2} \sum_{j,j'} \gamma^{(1)}_{\mu}(j,j')
 \langle \psi^*_j(0) \sigma^a \psi_{j'}(0) \rangle , \\
 \label{eq: def Jab}
 \cJ_\munu^{ab} &=& - \frac{1}{4} \sum_{j,j'} \sum_{l,l'}
 \gamma^{(1)}_{\mu}(j,j') \gamma^{(1)}_{\nu}(l,l')
 \nonumber \\
 && \times \int_0^\beta d\tau \,
 \big\langle \left( \psi^*_j(\tau) \sigma^a \psi_{j'}(\tau) \right)
 \left( \psi^*_l(0) \sigma^b \psi_{l'}(0) \right) \big\rangle_c
 \nonumber \\
 &+& \frac{1}{4} \sum_{j,j'} \gamma^{(2)}_{\mu\nu}(j,j') 
 \langle \psi^*_j(0) \psi_{j'}(0) \rangle \, \delta_{ab} ,
\label{eq: spin stiff definitions}
\end{eqnarray}
where $\langle\bullet\rangle$ ($\langle\bullet\rangle_c$) denotes the (connected) average with respect to the chargon Hubbard action. The first and second order current vertices have been defined as
\begin{subequations}
\begin{align}
\label{eq: gamma1}
 \gamma^{(1)}(j,j') =& \left(
 \delta_{jj'}, i\,x_{jj'} \, t_{jj'}, i\,y_{jj'} \, t_{jj'} \right) , \hskip 1cm \\
\label{eq: gamma2}
 \gamma^{(2)}(j,j') =& - \left( \begin{array}{ccc}
 0 & 0 & 0 \\
 0 & x_{jj'} x_{jj'} \, t_{jj'} & x_{jj'} y_{jj'} \, t_{jj'}\\
 0 & y_{jj'} x_{jj'} \, t_{jj'} & y_{jj'} y_{jj'} \, t_{jj'}\\
 \end{array} \right) , \hskip -5mm
\end{align}
\end{subequations}
where $x_{jj'}$ and $y_{jj'}$ are the $x$ and $y$ components, respectively of $\br_{jj'} = \br_j - \br_{j'}$.

In Appendix~\ref{app: linear term} we will see that the linear term in~\eqref{eq: effective action Amu} vanishes. We therefore consider only the quadratic contribution to the effective action. Defining the adjoint representation $\cR$ of the SU(2) rotation $R$ via
\begin{equation}
 R^\dagger \, \sigma^\a R = \cR^{ab} \sigma^\b ,
\label{eq: R to mathcal R}
\end{equation}
we obtain the non-linear sigma model (NL$\sigma$M) action for the directional fluctuations (see Appendix~\ref{app: derivation of the NLsM})
\begin{equation}
 \mathcal{S}_\mathrm{NL\sigma M}[\cR] = \int_\mathcal{T}\!dx \,
 \frac{1}{2}\Tr\left[\cP_\munu (\dmu\cR^T)(\dnu\cR)\right],
\label{eq: general NLsM}
\end{equation}
where $\cP_\munu = \frac{1}{2} \Tr[\cJ_\munu] \mathbb{1} - \cJ_\munu $.

The structure of the matrices $\cJ_\munu $ and $\cP_\munu $ depends on the magnetically ordered chargon state. In the trivial case $\langle \vec{S}^\psi_j \rangle = 0$ all the stiffnesses vanish and no meaningful low energy theory for $R$ can be derived. A well-defined low-energy theory emerges, for example, when \emph{N\'eel} antiferromagnetic order is realized in the chargon sector, that is,
\begin{equation}
 \langle \vec{S}^\psi_j \rangle \propto (-1)^{\boldsymbol{r}_j} \hat{u},
\end{equation}
where $\hat{u}$ is an arbitrary fixed unit vector. Choosing $\hat{u} = \hat{e}_1 = (1,0,0)$, the spin stiffness matrix in the N\'eel state has the form
\begin{equation}
 \cJ_\munu = \left( \begin{array}{ccc}
 0  & 0 & 0 \\ 0 & J_\munu  & 0 \\ 0 & 0 & J_\munu \end{array} \right) ,
\end{equation}
with $(J_{\mu\nu}) = {\rm diag}(-Z,J,J)$.
In this case the effective theory reduces to the well-known ${\rm O(3)/O(2)} \simeq S_2$ non-linear sigma model \cite{Haldane1983_I,Haldane1983_II}
\begin{equation}
 \mathcal{S}_\mathrm{NL\sigma M} =
 \frac{1}{2} \int_\mathcal {T} dx \, \left(
 Z |\partial_\tau\hat{\Omega}|^2 + J |\vec{\nabla}\hat{\Omega}|^2
 \right) ,
\end{equation}
where $\hat{\Omega}^\a=\cR^{\a1}$, and $|\hat{\Omega}|^2=1$.

Another possibility is planar \emph{spiral} magnetic ordering of the chargons,
\begin{equation}
 \langle\vec{S}^\psi_j\rangle \propto
 \cos(\bQ \cdot \br_j)\hat{u}_1 +
 \sin(\bQ \cdot \br_j)\hat{u}_2,
\end{equation}
where $\bQ$ is a fixed wave vector as obtained by minimizing the chargon free energy, while $\hat{u}_1$ and $\hat{u}_2$ are two arbitrary mutually orthogonal unit vectors. The special case $\bQ = (\pi,\pi)$ corresponds to the N\'eel state. Fixing $\hat{u}_1$ to $\hat{e}_1$ and $\hat{u}_2$ to $\hat{e}_2\equiv(0,1,0)$, the spin stiffness matrix assumes the form
\begin{equation}
  \cJ_\munu = \left( \begin{array}{ccc}
  J_\munu^\perp & 0 & 0 \\
  0 & J_\munu^\perp & 0 \\
  0 & 0 & J_\munu^\Box
  \end{array} \right),
\label{eq:spiral stiffness matrix}
\end{equation}
where
\begin{equation}
 (J_{\mu\nu}^a) =
 \left( \begin{array}{ccc}
 -Z^a & 0 & 0 \\  0 & J_{xx}^a & J_{xy}^a \\ 0 & J_{yx}^a & J_{yy}^a
 \end{array} \right) .
\end{equation}
for $a \in \{ \perp,\Box \}$.
In this case, the effective action maintains its general form~\eqref{eq: general NLsM} and it describes the O(3)$\times$O(2)/O(2) symmetric NL$\sigma$M, which has been previously studied in the context of geometrically frustrated antiferromagnets~\cite{Azaria1990,Azaria1992,Azaria1993_PRL,Azaria1993,Klee1996}. This theory has three independent degrees of freedom, corresponding to one \emph{in-plane} and two \emph{out-of-plane} Goldstone modes. 

In the following we will restrict the magnetic ordering pattern of the chargons to N\'eel or planar spiral order. N\'eel or spiral antiferromagnetism has been found in the two-dimensional Hubbard model over broad regions of the parameter space by several approximate methods, such as Hartree-Fock~\cite{Igoshev2010}, slave boson mean-field theory~\cite{Fresard1991}, expansion in the hole density~\cite{Chubukov1995}, moderate coupling fRG~\cite{Yamase2016}, and dynamical mean-field theory~\cite{Vilardi2018,Bonetti2020_I}. In our theory the mean-field order applies only to the chargons, while the physical electrons are subject to order parameter fluctuations.


\section{Computation of parameters}
\label{sec: fRG+MF}
In this section, we describe how we evaluate the chargon integral in Eq.~\eqref{eq: integral over psi} to compute the magnetic order parameter and the stiffness matrix $\cJ_{\mu\nu}$. The advantage of the way we formulated our theory in Sec.~\ref{sec: SU(2) gauge theory} is that it allows arbitrary approximations on the chargon action. One can employ various techniques to obtain the order parameter and the spin stiffnesses in the magnetically ordered phase. We use a renormalized mean-field (MF) approach with effective interactions obtained from a functional renormalization group (fRG) flow. In the following we briefly describe our approximation of the (exact) fRG flow, and we refer to Refs.~\cite{Berges2002, Metzner2012, Dupuis2021} for the fRG, and to Refs.~\cite{Wang2014, Yamase2016, Bonetti2020_II, Vilardi2020} for the fRG+MF method. 


\subsection{Symmetric regime}
We evaluate the chargon functional integral by using an fRG flow equation \cite{Berges2002, Metzner2012, Dupuis2021}, choosing the temperature $T$ as flow parameter \cite{Honerkamp2001}. Temperature can be used as a flow parameter after rescaling the chargon fields as $\psi_j \to T^\frac{3}{4}\psi_j$, and defining a rescaled bare Green's function,
$G_0^T(\bk,i\nu_n) = T^{\frac{1}{2}}/(i\nu_n - \epsilon_\bk + \mu)$,
where $\nu_n = (2n+1)\pi T$ is the fermionic Matsubara frequency, and $\epsilon_\bk$ is the Fourier transform of the hopping matrix in~\eqref{eq: Hubbard action}.

We approximate the exact fRG flow by a second order (one-loop) flow of the two-particle vertex $V^T$, discarding self-energy feedback and contributions from the three-particle vertex \cite{Metzner2012}. In a SU(2) invariant system the two-particle vertex has the spin structure
\begin{equation*}
\begin{split}
 V^T_{\sigma_1\sigma_2\sigma_3\sigma_4}(k_1,k_2,k_3,k_4) &=
 V^T(k_1,k_2,k_3,k_4) \, \delta_{\sigma_1\sigma_3}\,\delta_{\sigma_2\sigma_4} \\
 & - V^T(k_2,k_1,k_3,k_4) \, \delta_{\sigma_1\sigma_4}\,\delta_{\sigma_2\sigma_3} , 
\end{split}
\end{equation*}
where $k_\alpha = (\bk_\alpha,i\nu_{\alpha n})$ are combined momentum and frequency variables. Translation invariance imposes momentum conservation so that $k_1 + k_2 = k_3 + k_4$.
We perform a static approximation, that is, we neglect the frequency dependency of the vertex. To parametrize the momentum dependence, we use the channel decomposition~\cite{Husemann2009, Husemann2012, Vilardi2017, Vilardi2019}
\begin{eqnarray}
 && V^T(\bk_1,\bk_2,\bk_3,\bk_4) = U
 - \phi^{p,T}_{\frac{\bk_1-\bk_2}{2},\frac{\bk_3-\bk_4}{2}}(\bk_1+\bk_2) \nonumber \\
 && + \, \phi^{m,T}_{\frac{\bk_1+\bk_4}{2},\frac{\bk_2+\bk_3}{2}}(\bk_2-\bk_3)
   + \frac{1}{2}\phi^{m,T}_{\frac{\bk_1+\bk_3}{2},\frac{\bk_2+\bk_4}{2}}(\bk_3-\bk_1) 
   \nonumber \\
 && - \frac{1}{2}\phi^{c,T}_{\frac{\bk_1+\bk_3}{2},\frac{\bk_2+\bk_4}{2}}(\bk_3-\bk_1) ,
\label{eq: vertex parametrization}
\end{eqnarray}
where the functions $\phi^{p,T}$, $\phi^{m,T}$, and $\phi^{c,T}$ capture fluctuations in the pairing, magnetic, and charge channel, respectively.
The dependences of these functions on the linear combination of momenta in the brackets are typically much stronger than those in the subscripts. Hence, we expand the latter dependencies in form factors \cite{Husemann2009,Lichtenstein2017}, keeping only the lowest order s-wave, extended s-wave, p-wave and d-wave contributions.

We run the fRG flow from the initial temperature $T_\mathrm{ini} = \infty$, at which $V^{T_\mathrm{ini}} = U$, down to a critical temperature $T^*$ at $V^T$ diverges, signaling the onset of spontaneous symmetry breaking (SSB). If the divergence of the vertex is due to $\phi^{m,T}$, the chargons develop some kind of magnetic order. 


\subsection{Order parameter}
\label{sec: order parameter and Q}
In the magnetic phase, that is, for $T < T^*$, we assume an order parameter of the form
$\langle \psi^*_{\bk,\up} \psi_{\bk+\bQ,\down} \rangle$,  which corresponds to N\'eel antiferromagnetism if $\bQ = (\pi,\pi)$, and to spiral order otherwise. 

For $T < T^*$ we simplify the flow equations by decoupling the three channels $\phi^{p,T}$, $\phi^{m,T}$, and $\phi^{c,T}$. The flow equations can then be formally integrated, and the formation of an order parameter can be easily taken into account \cite{Wang2014}.
We focus on magnetic order and ignore the pairing instability to analyze the non-superconducting ``normal'' state. In the magnetic channel one thus obtains the magnetic gap equation ~\cite{Yamase2016}
\begin{equation}
 \Delta_{\bk} = \int_{\bk'} \bar{V}^m_{\bk,\bk'}(\bQ)\,
 \frac{f(E^-_{\bk'}) - f(E^+_{\bk'})}{E^+_{\bk'} - E^-_{\bk'}} \, \Delta_{\bk'} ,
\label{eq: gap equation fRG+MF}
\end{equation}
where $f(x)=(e^{x/T}+1)^{-1}$ is the Fermi function, $\int_\bk$ is a shorthand notation for $\int\!\frac{d^2\bk}{(2\pi)^2}$, and $E^\pm_\bk$ are the quasiparticle dispersions
\begin{equation}
 E^\pm_\bk = \frac{\epsilon_\bk+\epsilon_{\bk+\bQ}}{2}
 \pm\sqrt{\frac{1}{4} \left( \epsilon_\bk-\epsilon_{\bk+\bQ} \right)^2
 + \Delta_\bk^2} \, -\mu .
\end{equation}
The effective coupling $\bar{V}^m_{\bk,\bk'}(\bQ)$ is the particle-hole irreducible part of $V^{T^*}$ in the magnetic channel, which can be obtained by inverting a Bethe-Salpeter equation at the critical scale, 
\begin{equation}
\begin{split}
 V^{m,T^*}_{\bk,\bk'}(\bq) &= \bar{V}^m_{\bk,\bk'}(\bq) \\
 & - \int_{\bk''} \bar{V}^m_{\bk,\bk''}(\bq) \, \Pi^{T^*}_{\bk''}(\bq) \,
 V^{m,T^*}_{\bk'',\bk'}(\bq) ,
\end{split}
\label{eq: V phx Bethe-Salpeter}
\end{equation}
where $V^{m,T}_{\bk,\bk'}(\bq) = V^{T}(\bk-\bq/2,\bk'+\bq/2,\bk'-\bq/2,\bk+\bq/2)$, and the particle-hole bubble is given by
\begin{equation}
 \Pi^{T}_{\bk}(\bq) = \sum_{\nu_n} G_0^{T}\left(\bk-\bq/2,i\nu_n\right)
 G_0^{T}\left(\bk+\bq/2,i\nu_n\right) . 
\end{equation}
Although $V^{m,T^*}_{\bk,\bk'}(\bq)$ diverges at certain wave vectors $\bq = \bQ_c$, the irreducible coupling $\bar{V}^m_{\bk,\bk'}(\bq)$ is finite for all $\bq$.

The dependence of $\bar{V}^m_{\bk,\bk'}(\bq)$ on $\bk$ and $\bk'$ is rather weak and of no qualitative importance. Hence, to simplify the calculations, we discard the $\bk$ and $\bk'$ dependencies of the effective coupling by taking the momentum average
$\bar{V}^m(\bq) = \int_{\bk,\bk'} \bar{V}^m_{\bk,\bk'}(\bq)$.
The magnetic gap then becomes momentum independent, that is, $\Delta_\bk = \Delta$.
While the full vertex $V^{m,T}_{\bk,\bk'}(\bq)$ depends very strongly on $\bq$, the dependence of its irreducible part $\bar{V}_{\bk,\bk'}(\bq)$ on $\bq$ is rather weak. The calculation of the stiffnesses in the subsequent section is considerably simplified approximating $\bar{V}^m(\bq)$ by a momentum independent effective interaction $U_{\rm eff}^m = \bar{V}^m(\bQ_c)$.

The optimal ordering wave vector $\bQ$ is found by minimizing the mean-field free energy of the system
\begin{equation} \label{eq: MF theromdynamic potential}
 F(\bQ) = - T \int_\bk\sum_{\ell=\pm} \ln\left(1+e^{-E^\ell_\bk(\bQ)/T}\right)
 + \frac{\Delta^2}{2U_{\rm eff}^m} + \mu n ,
\end{equation}
where the chemical potential $\mu$ is determined by keeping the density
$n = \int_\bk \sum_{\ell=\pm} f(E^\ell_\bk)$ fixed. The optimal wave vectors $\bQ$ at temperatures $T < T^*$ generally differ from the wave vectors $\bQ_c$ at which $V^{T^*}_{\bk,\bk'}(\bq)$ diverges.

Eq.~\eqref{eq: gap equation fRG+MF} has the form of a mean-field gap equation with a renormalized interaction that is reduced compared to the bare Hubbard interaction $U$ by fluctuations in the pairing and charge channels. This reduces the critical doping beyond which magnetic order disappears, compared to the unrealistically large values obtained already for weak bare interactions in pure Hartree-Fock theory (see e.g.\ Ref.~\cite{Igoshev2010}). 


\subsection{Spin stiffnesses}
\label{sec: spin stiff formalism}

The NL$\sigma$M parameters, that is, the spin stiffnesses $\cJ_{\mu\nu}^{ab}$, are obtained by evaluating Eq.~\eqref{eq: spin stiff definitions}. These expressions can be viewed as the reponse of the chargon system to an external SU(2) gauge field in the low energy and long wavelength limit, and they are equivalent to the stiffnesses defined by an expansion of the inverse susceptibilities to quadratic order in momentum and frequency around the Goldstone poles \cite{Bonetti2022,Bonetti2022a}.
The following evaluation is obtained as a simple generalization of the RPA formula derived in Ref.~\cite{Bonetti2022a} to a renormalized RPA with effective interactions $U_{\rm eff}^m$ and $U_{\rm eff}^c$.
Since in the magnetic state the translational symmetry is broken, the Fourier transforms of the response functions depend on two distinct momenta $\bq$ and $\bq'$, where $\bq'$ can assume the values $\bq$, $\bq \pm \bQ$, and $\bq \pm 2\bQ$. However, to compute $\cJ_{\mu\nu}^{ab}$, we only need to deal with the limit $\bq,\bq' \to \mathbf{0}$. 

The temporal components of the stiffness matrix, that is, $\cJ_{00}^{ab}$, are given by the uniform spin susceptibility in the {\em dynamical}\/ limit \cite{Bonetti2022a}
\begin{equation}
 \cJ_{00}^{ab} = - \chi_{\rm dyn}^{ab} = - \lim_{\omega\to 0}
 \chi^{ab}(\mathbf{0},\mathbf{0},\omega) ,
\label{eq: temporal stiffness}
\end{equation}
where $\chi^{ab}(\bq,\bq',\omega)$ is the Fourier transform of
\begin{equation}
\chi_{jl}^{ab}(\tau) = \frac{1}{4} \big\langle \left( 
 \psi^*_j(\tau) \sigma^a \psi_j(\tau) \right)
 \left( \psi^*_l(0) \sigma^b \psi_l(0) \right)
 \big\rangle_c .
\label{eq: spin susceptibility}
\end{equation}
Note that, in a metallic system, the {\em static}\/ uniform susceptibility obtained from $\bq,\bq' \to \mathbf{0}$ after setting $\omega = 0$ differs from the quantity defined in Eq.~\eqref{eq: temporal stiffness}.

The spin susceptibility can be most conveniently computed in a rotating spin frame defined by the transformation \cite{Kampf1996,Bonetti2022}
\begin{equation} 
 \widetilde\psi_j = e^{-i\frac{\bQ}{2} \cdot \mathbf{r}_j}
 e^{i\sigma^3\frac{\bQ}{2} \cdot \mathbf{r}_j} \psi_j , \quad
 \widetilde{\psi}^*_j = \psi^*_j\, e^{-i\sigma^3\frac{\bQ}{2} \cdot \mathbf{r}_j}
 e^{i\frac{\bQ}{2} \cdot \mathbf{r}_j} ,
\label{eq: chargon rotation}
\end{equation}
since in the rotated frame the magnetically ordered system appears translation invariant. Hence, the rotated susceptibility is diagonal in momentum space and can therefore be written as $\chit^{ab}(\bq,\omega)$, with a single momentum variable $\bq$.

Consistently with the mean-field theory for the magnetic order parameter, we compute the susceptibilities in the magnetic state via a random phase approximation (RPA) with renormalized interactions as obtained from the fRG. In a spiral state with a generic wave vector $\bQ$, the spin susceptibility is coupled to the charge susceptibility \cite{Kampf1996}. Hence, we extend the definition of the spin susceptibility in Eq.~\eqref{eq: spin susceptibility} to a combined charge-spin susceptibility by including the value $0$ for the indices $a$ and $b$, in addition to the values $1,2,3$, and defining  $\sigma^0$ as the two-dimensional unit matrix.
The prefactor $\frac{1}{4}$ in Eq.~\eqref{eq: spin susceptibility} implies that $\chi^{00}$ is actually a quarter of the conventional charge susceptibility.
                                                                                                                                                                                                                                                                                                                                                                                                                                                                                                                                                                                                                                                                   
In RPA, the rotated susceptibility $\chit$ can be written as
\begin{equation}
 \chit^{ab}(q) = \chit_{0}^{ab}(q) +
 \sum_{a',b'=0}^3 \chit_{0}^{aa'}(q) \widetilde{\Gamma}^{a'b'}(q) \chit_{0}^{b'b}(q),
\label{eq: chitilde general expression}
\end{equation}
where $q = (\bq,\omega)$, and $\widetilde\Gamma^{ab}(q)$ is the RPA effective interaction in the rotated spin frame.
The ``bare'' susceptibility $\chit_{0}^{ab}(q)$ is given by the particle-hole bubble \cite{Bonetti2022}
\begin{eqnarray}
 \chit_{0}^{ab}(\bq,\omega) &=& - \frac{1}{4} \int_\bk T \sum_{\nu_n} 
 \Tr \big[ \sigma^a \widetilde{\mathcal{G}}(\bk+\bq,i\nu_n+\omega+i0^+) \nonumber \\
 && \times \, \sigma^b \widetilde{\mathcal{G}}(\bk,i\nu_n) \big] \, ,
\end{eqnarray}
with $\widetilde{\mathcal{G}}(\bk,i\nu_n)$ the mean-field chargon Green's function in the rotated basis
\begin{equation}
 \widetilde{\mathcal{G}}(\bk,i\nu_n) = 
 \left( \begin{array}{cc}
 i\nu_n - \epsilon_\bk + \mu & -\Delta \\
 -\Delta & i\nu_n - \epsilon_{\bk+\bQ} + \mu
 \end{array} \right)^{-1}.
\end{equation}
The RPA effective interaction is obtained from a ladder sum, leading to the linear matrix equation
\begin{equation} \label{eq: ladder eff interaction}
 \widetilde{\Gamma}^{ab}(q) = \widetilde{\Gamma}^{ab}_{0}(\bq) +
 \sum_{a',b'=0}^3 \widetilde{\Gamma}^{aa'}_{0}(\bq) \, \chit^{a'b'}_{0}(q) \,
 \widetilde{\Gamma}^{b'b}(q),
\end{equation}
where
\begin{equation}
 \widetilde{\Gamma}^{ab}_{0}(\bq) = \Gamma^{ab}_{0}(\bq) =
 2 \, {\rm diag} \left[
 -U_{\rm eff}^c(\bq),U_{\rm eff}^m,U_{\rm eff}^m,U_{\rm eff}^m \right] .
\end{equation}
The effective charge interaction is given by
$U_{\rm eff}^c(\bq) = \int_{\bk,\bk'} \bar V^c_{\bk,\bk'}(\bq)$,
where the irreducible coupling $\bar V^c_{\bk,\bk'}(\bq)$ is obtained by inverting a Bethe-Salpeter equation similar to Eq.~\eqref{eq: V phx Bethe-Salpeter},
\begin{equation}
\begin{split}
 V^{c,T^*}_{\bk,\bk'}(\bq) &= \bar{V}^c_{\bk,\bk'}(\bq) \\
 & + \int_{\bk''} \bar{V}^c_{\bk,\bk''}(\bq) \, \Pi^{T^*}_{\bk''}(\bq) \,
 V^{c,T^*}_{\bk'',\bk'}(\bq) ,
\end{split}
\label{eq: V ph Bethe-Salpeter}
\end{equation}
with
\begin{eqnarray}
 V^{c,T}_{\bk,\bk'}(\bq) &=& 2V^{T}(\bk\!-\!\bq/2,\bk'\!+\!\bq/2,\bk\!+\!\bq/2,\bk'\!-\!\bq/2)
 \nonumber \\
 &-& V^{T}(\bk\!-\!\bq/2,\bk'\!+\!\bq/2,\bk'\!-\!\bq/2,\bk\!+\!\bq/2) . \nonumber
\end{eqnarray}
Here we keep the dependence on $\bq$ since it does not complicate the calculations.  
The off-diagonal ($a \neq b$) elements of $\chi^{ab}(\mathbf{0},\mathbf{0},\omega)$ with $a,b = 1,2,3$ vanish for $\omega \to 0$ both in the spiral and in the N\'eel state, so that we need to deal only with the diagonal spin susceptibility components $\chi^{aa}(\bq,\bq,\omega)$.

In a spiral state with $\bQ \neq (\pi,\pi)$, the diagonal (both in momentum and spin indices) spin susceptibility components are related to the susceptibility components in the rotated basis as \cite{Bonetti2022}
\begin{subequations} \label{eq: chi phys from chi tilde}
 \begin{align}
 \chi^{11}(\bq,\bq,\omega) &= \chi^{22}(\bq,\bq,\omega) \nonumber \\
 &= \frac{1}{4} \sum_{s=\pm} \big[ \chit^{11}(q+sQ) + \chit^{22}(q+sQ) \big] \nonumber \\ 
 & \quad + 2i\chit^{12}(q+Q) + 2i\chit^{21}(q-Q) , \\
 \chi^{33}(\bq,\bq,\omega) &= \chit^{33}(q) ,
 \end{align}
\end{subequations}
where $Q=(\bQ,0)$.
The momentum diagonal components of $\chi^{11}(\bq,\bq',\omega)$ and $\chi^{22}(\bq,\bq',\omega)$ are equal, and the limit $\omega\to 0$ in \eqref{eq: temporal stiffness} is nonzero for all diagonal components, yielding
\begin{equation}
 \cJ_{00} = \left(
 \begin{array}{ccc}
 - Z^\perp & 0 & 0 \\ 0 & - Z^\perp & 0 \\ 0 & 0 & - Z^\Box
 \end{array} \right),
\end{equation}
with $Z^\perp = \chi^{22}(\mathbf{0},\mathbf{0},\omega \to 0)$ and
$Z^\Box = \chi^{33}(\mathbf{0},\mathbf{0},\omega\to0)$.
The quantities $Z^\Box$ and $Z^\perp$ parametrize the low frequency dependence of the in-plane and out-of-plane spin susceptibility, respectively, near the Goldstone poles \cite{Bonetti2022, Bonetti2022a}.

In the limits $\omega \to 0$ and $\bq \to \mathbf{0}$ or $\bQ$, several off-diagonal matrix elements of the RPA effective interaction $\widetilde\Gamma^{ab}(\bq,\omega)$ vanish \cite{Bonetti2022}. The expressions for $Z^a$ can therefore be simplified to \cite{Bonetti2022a}
\begin{equation} \label{eq: ZBox}
 Z^\Box = \frac{\chit_0^{33}(\mathbf{0},\omega \to 0)}
 {1 - 2U_{\rm eff}^m \, \chit_0^{33}(\mathbf{0},\omega \to 0)} \, ,
\end{equation}
and
\begin{eqnarray} \label{eq: Zperp}
 Z^\perp &=& 2 \chit_0^{-+}(\bQ,0) \nonumber \\
 &+& 2 \sum_{a,b = 0,1,2}
 \chit_0^{-a}(\bQ,0) \widetilde\Gamma^{ab}(\bQ,0) \chit_0^{b+}(\bQ,0) \, , \hskip 5mm
\end{eqnarray}
where superscripts $+$ and $-$ attached to $\chit_0^{\phantom{a}}$ indicate that the susceptibilities are formed with the ladder operators $S^{\pm} = \frac{1}{2} \left( S^1 \pm S^2 \right)$.

In the N\'eel state, terms which contribute to $\chi^{11}(\bq,\bq',\omega)$ and $\chi^{22}(\bq,\bq',\omega)$  with $\bq' = \bq \pm 2\bQ$ for $\bQ \neq (\pi,\pi)$, contribute to the momentum diagonal susceptibilities since $2\bQ \equiv \mathbf{0}$ for $\bQ = (\pi,\pi)$. The transformation of the susceptibilities from the rotated to the unrotated basis then reads \cite{Bonetti2022}
\begin{subequations} \label{eq: chi phys from chi tilde Neel}
 \begin{align}
 &\chi^{11}(\bq,\bq,\omega)=\chit^{11}(q+Q),\\
 &\chi^{22}(\bq,\bq,\omega)=\chit^{22}(q+Q),\\
 &\chi^{33}(\bq,\bq,\omega)=\chit^{33}(q).
 \end{align}
\end{subequations}
Since $\chit^{11}(\bQ,\omega)=0$ and $\chit^{22}(q+Q) = \chit^{33}(q)$ for $\bQ = (\pi,\pi)$, we obtain
\begin{equation}
 \cJ_{00} = \left(
 \begin{array}{ccc}
 0 & 0 & 0 \\ 0 & -Z & 0 \\ 0 & 0 & -Z
 \end{array} \right),
\end{equation}
with $Z = \chi^{22}(\mathbf{0},\mathbf{0},\omega\to 0)$ in the N\'eel state, which can be evaluated by the same expression as the one for $Z^\Box$ in Eq.~\eqref{eq: ZBox}.


%
\begin{figure*}[t]
    \centering
    \includegraphics[width=1.\textwidth]{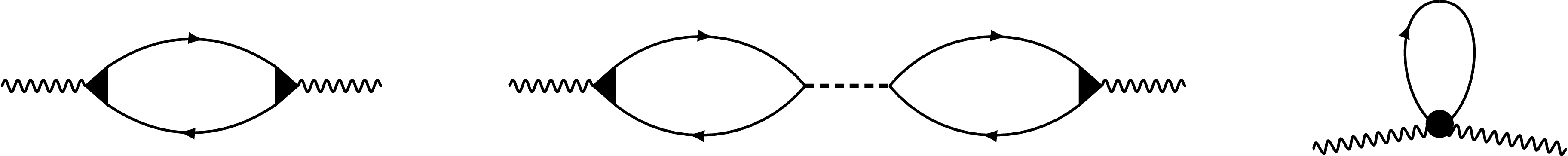}
    \caption{Diagrams contributing to the spin stiffness. The black triangles and circles represent the first and second order current vertices $\gamma_\alpha^{(1)}(\bk)$ and $\gamma_{\alpha\beta}^{(2)}(\bk)$, respectively, while the dashed line represents the effective interaction $\Gamma^{ab}(\bq,\bq',\omega)$. }
    \label{fig: Fig1}
\end{figure*}
The spatial components of the stiffness matrix $\cJ_{\alpha\beta}^{ab}$ with $\alpha,\beta = 1,2$ are obtained from the spatial components of the uniform gauge field kernel $\cK_{\alpha\beta}^{ab}$ in the {\em static}\/ limit \cite{Bonetti2022a}
\begin{equation}
 \cJ_{\alpha\beta}^{ab} =
 - \lim_{\bq \to \mathbf{0}} \cK_{\alpha\beta}^{ab}(\bq,\bq,0) \, ,
\end{equation}
where $\cK_{\alpha\beta}^{ab}(\bq,\bq',\omega) =
\cK_{\alpha\beta}^{{\rm p},ab}(\bq,\bq',\omega) + \delta_{ab} \cK_{\alpha\beta}^{\rm d}$
is the Fourier transform of
\begin{equation}
 \cK_{\alpha\beta,jl}^{ab}(\tau) = \cK_{\alpha\beta,jl}^{{\rm p},ab}(\tau) +
 \cK_{\alpha\beta}^{\rm d} \, \delta_{ab} \delta_{jl} \delta(\tau) ,
\end{equation}
with the paramagnetic and diamagnetic contributions (cf.\ Eq.~\eqref{eq: def Jab})
\begin{eqnarray}
 \label{eq: defKpara}
 \cK_{\alpha\beta,jl}^{{\rm p},ab}(\tau) &=&
 \frac{1}{4} \sum_{j',l'} \gamma_{\alpha}^{(1)}(j,j') \, \gamma_{\beta}^{(1)}(l,l')
 \nonumber \\
 \label{eq: defKdia}
 &\times& \big\langle \left( \psi^*_j(\tau) \sigma^a \psi_{j'}(\tau) \right)
 \left( \psi^*_l(0) \sigma^b \psi_{l'}(0) \right) \big\rangle_c \, , \hskip 6mm \\[2mm]
 \cK_{\alpha\beta}^{\rm d} &=&
 - \frac{1}{4} \sum_{j'} \gamma_{\alpha\beta}^{(2)}(j,j') \,
 \big\langle \psi^*_j(0) \psi_{j'}(0) \big\rangle \, ,
\end{eqnarray}
respectively. The diamagnetic contribution is translation invariant (in a spiral or N\'eel state). Fourier transforming, and evaluating the expectation value in Eq.~\eqref{eq: defKpara} by the renormalized RPA, the paramagnetic part of the spin stiffness can be written as
\begin{eqnarray} \label{eq: Jp general expression}
 \cJ_{\alpha\beta}^{{\rm p},ab} &=& - \lim_{\bq\to\mathbf{0}} \Big[
 \cK_{0,\alpha\beta}^{{\rm p},ab}(\bq,\bq,0) \nonumber \\
 &+& \sum_{a',b'=0}^3 \int_{\bq',\bq''}
 \cK_{0,\alpha 0}^{{\rm p},aa'}(\bq,\bq',0) \nonumber \\
 && \times \, \Gamma^{a'b'}(\bq',\bq'',0) \,
 \cK_{0,0\beta}^{{\rm p},b'b}(\bq'',\bq,0) \Big] ,
\end{eqnarray}
where $\Gamma^{ab}(\bq,\bq',\omega)$ is the RPA effective interaction~\eqref{eq: ladder eff interaction} in the original (non-rotated) spin basis. The bare paramagnetic response kernel is given by
\begin{eqnarray}
 \cK_{0,\mu\nu}^{{\rm p},ab}(\bq,\bq',0) &=&
 - \frac{1}{4} \int_{\bk,\bk'} T \sum_{\nu_n}
 \gamma_\mu^{(1)}(\bk+\bq) \, \gamma_\nu^{(1)}(\bk') \nonumber \\
 && \hskip -2cm \times \, \Tr\left[ \sigma^a \mathcal{G}(\bk+\bq,\bk'+\bq',i\nu_n)
 \sigma^b \mathcal{G}(\bk',\bk,i\nu_n)\right] , \hskip 7mm
\end{eqnarray}
where $\gamma^{(1)}(\bk) = (1,\partial_{k_x}\epsilon_\bk,\partial_{k_y}\epsilon_\bk)$ is the Fourier transform of $\gamma^{(1)}(j,j')$ in Eq.~\eqref{eq: gamma1}.
The chargon Green's function in the original spin basis reads
\begin{equation} \label{eq: calG}
 \mathcal{G}(\bk,\bk',i\nu_n) = \left(
 \begin{array}{cc}
 G_{+\bQ}(k) \, \delta_{\bk,\bk'} & F_{+\bQ}(k) \, \delta_{\bk+\bQ,\bk'} \\
 F_{-\bQ}(k) \, \delta_{\bk-\bQ,\bk'} & G_{-\bQ}(k) \, \delta_{\bk,\bk'}
 \end{array}\right) ,
\end{equation}
with 
\begin{subequations}
\begin{align}
 & G_{\pm\bQ}(k) = \frac{i\nu_n-\epsilon_{\bk\pm\bQ}+\mu}
 {(i\nu_n\!-\!\epsilon_{\bk}\!+\!\mu)(i\nu_n\!-\!\epsilon_{\bk\pm\bQ}\!+\!\mu) - \Delta^2} , \\
 & F_{\pm\bQ}(k) = \frac{\Delta}
 {(i\nu_n\!-\!\epsilon_{\bk}\!+\!\mu)(i\nu_n\!-\!\epsilon_{\bk\pm\bQ}\!+\!\mu) - \Delta^2} .
\end{align}
\end{subequations}

The diamagnetic part of the spin stiffness can be obtained from the Green's function as
\begin{equation}
 \cJ_{\alpha\beta}^{{\rm d},ab} =
 \frac{\delta_{ab}}{4} \int_{\bk,\bk'} T \sum_{\nu_n}
 \gamma_{\alpha\beta}^{(2)}(\bk) \, \tr \left[ \cG(\bk,\bk',i\nu_n) \right] \,
\end{equation}
where $\gamma_{\alpha\beta}^{(2)}(\bk) = \partial_{k_\alpha} \partial_{k_\beta} \epsilon_\bk$ is the Fourier transform of the second order current vertex $\gamma_{\alpha\beta}^{(2)}(j,j')$ in Eq.~\eqref{eq: gamma2}.
The various contributions to the response kernel are represented diagrammatically in Fig.~\ref{fig: Fig1}.

The off-diagonal ($a \neq b$) components of $\cK_{\alpha\beta}^{ab}(\bq,\bq,0)$ vanish for $\bq \to \mathbf{0}$ in the spiral and in the N\'eel state, so that we need to consider only the diagonal components.
For $a=b=1,2$ only the first (bare) term in Eq.~\eqref{eq: Jp general expression} contributes to the stiffness \cite{Bonetti2022a}. In a spiral state with $\bQ \neq (\pi,\pi)$, one thus obtains the out-of-plane stiffness as
\begin{equation} \label{eq: Jperp}
 J_{\alpha\beta}^\perp = \cJ_{\alpha\beta}^{11} = \cJ_{\alpha\beta}^{22}
 = - \lim_{\bq \to \mathbf{0}} \cK_{0,\alpha\beta}^{{\rm p},22}(\bq,\bq,0)
 - \cK_{\alpha\beta}^{\rm d} \, .
\end{equation}
For $a=b=3$, there are non vanishing components of the kernel that mix temporal and spatial indices, namely
\begin{subequations}
\begin{align}
 \cK_{0,\alpha 0}^{{\rm p},30}(\bq,\bq',0) &= \cK_{0,0 \alpha}^{{\rm p},03}(\bq,\bq',0) =
 \cK_{0,\alpha 0}^{{\rm p},30}(\bq,0) \, \delta_{\bq,\bq'} \, , \\
 \cK_{0,\alpha 0}^{{\rm p},31}(\bq,\bq',0) &= \cK_{0,0 \alpha}^{{\rm p},13}(\bq,\bq',0)
 \nonumber \\
 &= \cK_{0,\alpha 0}^{{\rm p},31}(\bq,0) \,
 \frac{\delta_{\bq+\bQ,\bq'} + \delta_{\bq-\bQ,\bq'}}{2} \, , \\
 \cK_{0,\alpha 0}^{{\rm p},32}(\bq,\bq',0) &= \cK_{0,0 \alpha}^{{\rm p},23}(\bq,\bq',0)
 \nonumber \\
 &= \cK_{0,\alpha 0}^{{\rm p},32}(\bq,0) \,
 \frac{\delta_{\bq+\bQ,\bq'} - \delta_{\bq-\bQ,\bq'}}{2i} \, .
\end{align}
\end{subequations}
Using $\cK_{0,\alpha 0}^{{\rm p},31}(\bq \to \mathbf{0},0) =
\cK_{0,\alpha 0}^{{\rm p},32}(\bq \to \mathbf{0},0)$, one thus obtains the in-plane stiffness in the form \cite{Bonetti2022a}
\begin{eqnarray} \label{eq: JBox}
 J_{\alpha\beta}^\Box &=& \cJ_{\alpha\beta}^{33} = 
 - \lim_{\bq \to \mathbf{0}} \cK_{0,\alpha\beta}^{{\rm p},33}(\bq,\bq,0)
 - \cK_{\alpha\beta}^{\rm d} \nonumber \\
 &-& \lim_{\bq \to 0} \sum_{a,b = 0,1} \cK_{0,\alpha 0}^{{\rm p},3a}(\bq,0) \,
 \widetilde\Gamma^{ab}(\bq,0) \, \cK_{0,\beta 0}^{{\rm p},3b}(\bq,0) \, , \hskip 7mm
\end{eqnarray}
where $\widetilde{\Gamma}^{ab}(q)$ is the effective interaction in the spin rotated basis, see Eq.~\eqref{eq: ladder eff interaction}.

In the N\'eel state one finds, in close analogy to the temporal components $\cJ_{00}^{ab}$ of the stiffness matrix, $\cJ_{\alpha\beta}^{11} = 0$ and $\cJ_{\alpha\beta}^{22} = \cJ_{\alpha\beta}^{33} = J \delta_{\alpha\beta}$, which can be most easily computed from the right hand side of Eq.~\eqref{eq: Jperp}.

In our low energy theory of the spinons we have ignored possible imaginary contributions from Landau damping of the Goldstone modes. In a N\'eel state, they are of the same order in the gradient expansion as the (real) temporal and spatial stiffness terms \cite{Sachdev1995}. The same is true for the Landau damping of the in-plane mode in a spiral state, but the damping of the out-of-plane mode is of higher order \cite{Bonetti2022}. Moreover, it requires the existence of hot spots (connected by $\bQ$) of the reconstructed Fermi surface. In our large $N$ evaluation of the NL$\sigma$M, the in-plane modes of the spiral state do not contribute. Hence, for the spiral state, Landau damping is irrelevant for our theory, while in the N\'eel state their might be a quantitative (not qualitative) modification of our results.

We conclude this section by comparing our theory to the SU(2) gauge theory of the half-filled Hubbard model derived by Borejsza and Dupuis \cite{Borejsza2004}. They used the same fractionalization of the electron in chargons and spinons, and the chargon order was treated in (plain) mean-field theory. Our expressions for the spin stiffnesses agree with theirs (at half-filling) if we replace our renormalized interaction $U_{\rm eff}^m$ by the bare Hubbard interaction $U$, although their derivation differs from ours. Following earlier work by Haldane \cite{Haldane1983_I, Haldane1983_II} for the Heisenberg model, Borejsza and Dupuis obtained their expressions for the spin stiffnesses by splitting the spinon into a ``N\'eel field'' and a ``canting field'' describing ferromagnetic fluctuations. Integrating out the fermions and the canting field they obtained a NL$\sigma$M for the N\'eel field, where the stiffnesses are given by the RPA. We obtain the same expressions (with a renormalized coupling) more directly from the RPA evaluation of the gauge field response, without introducing the canting field.

%
\section{Evaluation of sigma model}

To solve the NL$\sigma$M, we resort to a saddle point approximation in the $\text{CP}^{N-1}$ representation, which is exact in the large $N$ limit \cite{AuerbachBook1994,Chubukov1994}.


\subsection{\texorpdfstring{CP$^{\bf 1}$}{CP} representation}

The matrix $\cR$ can be expressed as a triad of orthonormal unit vectors:
\begin{equation} \label{eq: mathcal R to Omegas}
 \cR=\big( \hat\Omega_1,\hat\Omega_2,\hat\Omega_3 \big),
\end{equation}
where $\hat\Omega_i\cdot\hat\Omega_j = \delta_{ij}$. We represent these vectors in terms of two complex Schwinger bosons $z_\up$ and $z_\down$ \cite{Sachdev1995}
\begin{subequations} \label{eq: mathcal R to z}
\begin{align}
 & \hat\Omega_- = z(i\sigma^2\vec{\sigma})z, \\
 & \hat\Omega_+ = z^*(i\sigma^2\vec{\sigma})^\dagger z^*, \\
 & \hat\Omega_3 = z^*\vec{\sigma}z,
\end{align}
\end{subequations}
with $z = (z_\up,z_\down)$ and $\hat{\Omega}_\pm=\hat{\Omega}_1\mp i \hat{\Omega}_2$.
The Schwinger bosons obey the nonlinear constraint
\begin{equation} \label{eq: z boson constraint}
 z^*_\up z_\up + z^*_\down z_\down = 1 \, .
\end{equation}
The parametrization~\eqref{eq: mathcal R to z} is equivalent to
\begin{equation} \label{eq: R to z}
 R = \left( \begin{array}{cc}
 z_\up &  -z_\down^* \\ z_\down & \phantom{-} z_\up^* 
 \end{array} \right).
\end{equation}
Inserting the expressions \eqref{eq: mathcal R to Omegas} and \eqref{eq: mathcal R to z} into Eq.~\eqref{eq: general NLsM} and assuming a stiffness matrix $\cJ_\munu $ of the form~\eqref{eq:spiral stiffness matrix}, we obtain the $\rm CP^1$ action for fluctuating spiral order
\begin{eqnarray} \label{eq: CP1 action}
 \mathcal{S}_{\text{CP}^1}[z,z^*] &=& \int_\mathcal {T} dx \,
 \Big[ 2J^\perp_{\mu\nu} (\dmu z^*)(\dnu z) \nonumber \\
 && - \, 2(J^\perp_{\mu\nu} - J^\Box_{\mu\nu}) j_\mu j_\nu \Big] \, ,
\end{eqnarray}
with sum convention for repeated greek indices and the current operator
\begin{equation}
 j_\mu = \frac{i}{2}\left[z^*(\dmu z)-(\dmu z^*)z\right] \, .
\end{equation}
For the N\'eel case, the $\rm CP^1$ action is given by the same expression with $J^\Box_{\mu\nu} = 0$.
We recall that $x = (\tau,\br)$ comprises the imaginary time and space variables, and
$\mathcal{T} = [0,\beta] \times \mathbb{R}^2$.


\subsection{Large {\em N} expansion}

The current-current interaction in Eq.~\eqref{eq: CP1 action} can be decoupled by a Hubbard-Stratonovich transformation, introducing a U(1) gauge field $\mathcal{A}_\mu$, and implementing the constraint~\eqref{eq: z boson constraint} by means of a Lagrange multiplier $\lambda$. The resulting form of the action describes the so-called massive $\rm CP^1$ model~\cite{Azaria1995}
\begin{eqnarray} \label{eq: massive CP1 model}
 \cS_{\text{CP}^1}[z,z^*,\mathcal{A}_\mu,\lambda] &=& \int_\mathcal {T} dx
 \Big[ 2J^\perp_\munu (D_\mu z)^* (D_\nu z) \nonumber \\
 &+& \frac{1}{2} M_{\mu\nu} \mathcal{A}_\mu \mathcal{A}_\nu + i\lambda(z^*z-1)
 \Big] \, , \hskip 5mm
\end{eqnarray}
where $D_\mu = \dmu - i\mathcal{A}_\mu$ is the covariant derivative. The numbers $M_{\mu\nu}$ are the matrix elements of the mass tensor of the U(1) gauge field,
\begin{equation}
 {\rm M} = 4 \big[ 1 - {\rm J}^\Box ({\rm J}^\perp)^{-1} \big]^{-1} {\rm J}^\Box \, ,
\end{equation}
where ${\rm J}^\Box$ and ${\rm J}^\perp$ are the stiffness tensors built from the matrix elements
$J_{\mu\nu}^\Box$ and $J_{\mu\nu}^\perp$, respectively.

To perform a large $N$ expansion, we extend the two-component field $z = (z_\up,z_\down)$ to an $N$-component field $z = (z_1,\dots,z_N)$, and rescale it by a factor $\sqrt{N/2}$ so that it now satisfies the constraint
\begin{equation}
 z^*z = \sum_{\alpha=1}^N z^*_\alpha z_\alpha = \frac{N}{2} \, .
\end{equation}
To obtain a nontrivial limit $N \to \infty$, we rescale the stiffnesses $J^\perp_{\mu\nu}$ and $J^\Box_{\mu\nu}$ by a factor $2/N$, yielding the action
\begin{eqnarray} \label{eq: massive CPN1 model}
 \cS_{\text{CP}^{N-1}}[z,z^*,\mathcal{A}_\mu,\lambda] &=& \int_\mathcal {T} dx
 \Big[ 2J^\perp_\munu (D_\mu z)^* (D_\nu z) \nonumber \\
 &+& \! \frac{N}{4} M_{\mu\nu} \mathcal{A}_\mu \mathcal{A}_\nu +
 i\lambda \Big( z^*z - \frac{N}{2} \Big)
 \Big] . \hskip 7mm
\end{eqnarray}
This action describes the massive ${\rm CP}^{N-1}$ model~\cite{Campostrini1993}, which in $d>2$ dimensions displays two distinct critical points~\cite{Azaria1995,Chubukov1994,Chubukov1994_II}. The first one belongs to the pure ${\rm CP}^{N-1}$ class, where $M_{\mu\nu} \to 0$ ($J^\Box_{\mu\nu} = 0$), which applies, for example, in the case of N\'eel ordering of the chargons, and the U(1) gauge invariance is preserved. The second is in the O(2N) class, where $M_{\mu\nu} \to \infty$ ($J^\perp_{\mu\nu} = J^\Box_{\mu\nu}$) and the gauge field does not propagate. At the leading order in $N^{-1}$, the saddle point equations are the same for both fixed points, so that we can ignore this distinction in the following. 

At finite temperatures $T > 0$ the non-linear sigma model does not allow for any long-range magnetic order, in agreement with the Mermin-Wagner theorem. The spin correlations decay exponentially and the spin excitations are bounded from below by a spin gap
$m_s = \sqrt{i\langle\lambda\rangle/Z^\perp}$.
In the large $N$ limit, the spin gap $m_s$ is related to the spin stiffness by the following equation (see Appendix \ref{app: large N} for a derivation) 
\begin{equation} \label{eq: large N equation}
 \frac{1}{4\pi J}
 \int_0^{c_s\Luv} \!\frac{\epsilon\,d\epsilon}{\sqrt{\epsilon^2+m_s^2}} \,\mathrm{coth}\left(\frac{\sqrt{\epsilon^2+m_s^2}}{2T}\right) = 1 \, ,
\end{equation}
where $\Luv$ is an ultraviolet momentum cutoff. The constant $J$ is an ``average'' spin stiffness given by
\begin{equation}
 J = \sqrt{ \mathrm{det} \left( \begin{array}{cc}
 J^\perp_{xx} & J^\perp_{xy} \\ J^\perp_{yx} & J^\perp_{yy}
 \end{array} \right) } \, ,
\end{equation}
and $c_s = \sqrt{J/Z^\perp}$ is the corresponding average spin wave velocity.
In Sec.~\ref{sec: cutoff}, we shall discuss how to choose the value of $\Luv$. 
For $m_s \ll c_s\Luv$, and $T \ll c_s\Luv$, the magnetic correlation length $\xi_s = \frac{1}{2} c_s/m_s$, behaves as
\begin{equation}
 \xi_s = 
 \frac{c_s}{4T \, \sinh^{-1} \! \left[
 \frac{1}{2} e^{-\frac{2\pi}{T}(J - J_c)} \right] } \, ,
\end{equation}
with the critical stiffness
\begin{equation} \label{eq: Jc}
 J_c = \frac{c_s\Luv}{4\pi} \, .
\end{equation}
The correlation length is finite at each $T > 0$. For $J > J_c$, $\xi_s$ diverges exponentially for $T \to 0$, while for $J < J_c$ it remains finite in the zero temperature limit.

At $T=0$, the bosons may condense and the saddle point condition yields
\begin{equation} \label{eq: large N eq at T=0}
 n_0 + \frac{1}{4\pi J}
 \int_0^{c_s\Luv} \! \frac{\epsilon \, d\epsilon}{\sqrt{\epsilon^2 + m_s^2}} = 1 \, ,
\end{equation}
where $n_0 = |\langle z_1 \rangle|^2$ is the fraction of condensed bosons. Eq.~\eqref{eq: large N eq at T=0} can be easily solved, yielding (if $m_s \ll \Luv$)
\begin{subequations}
\begin{align}
 &\begin{cases}
 & m_s=0 \\ & n_0 = 1 - \frac{J_c}{J}
 \end{cases}
 \hskip 5mm \text{for } J>J_c \, , \\ 
 &\begin{cases}
 & n_0 = 0 \\ & m_s = 2\pi J\left[\left( J_c/J \right)^2 - 1 \right]
 \end{cases}
 \hskip 5mm \text{for } J<J_c.
\end{align}
\end{subequations}

The Mermin-Wagner theorem is thus respected already in the saddle-point approximation to the ${\rm CP}^{N-1}$ representation of the nonlinear sigma model, that is, there is no long-range order at $T > 0$. In the ground state, long-range order (corresponding to a $z$ boson condensation) is obtained for a sufficiently large spin stiffness, while for $J < J_c$ magnetic order is destroyed by quantum fluctuations even at $T = 0$, giving rise to a paramagnetic state with a spin gap.


\subsection{Choice of ultraviolet cutoff}
\label{sec: cutoff}
The impact of spin fluctuations described by the nonlinear sigma model depends strongly on the ultraviolet cutoff $\Luv$. In particular, the critical stiffness $J_c$ separating a ground state with magnetic long-range order from a disordered ground state is directly proportional to $\Luv$. The need for a regularization of the theory by an ultraviolet cutoff is a consequence of the gradient expansion. While the expansion coefficients (the stiffnesses) are determined by the microscopic model, there is no systematic way of computing $\Luv$.

A pragmatic choice for the cutoff is given by the ansatz
\begin{equation} \label{eq: Luv}
 \Luv = C/\xi_A \, ,
\end{equation}
where $C$ is a dimensionless number, and $\xi_A$ is the magnetic coherence length, which is the characteristic length scale of spin amplitude correlations. This choice may be motivated by the observation that local moments with a well defined spin amplitude are not defined at length scales below $\xi_A$ \cite{Borejsza2004}.
The constant $C$ can be fixed by matching results from the nonlinear sigma model to results from a microscopic calculation in a suitable special case (see below).

The coherence length $\xi_A$ can be obtained from the connected spin amplitude correlation function $\chi_A(\mathbf{r}_{j},\mathbf{r}_{j'}) = \big\langle (\hat{n}_j \cdot \vec{S}^\psi_j)(\hat{n}_{j'} \cdot \vec{S}^\psi_{j'}) \big\rangle_c$, where
$\hat{n}_j = \langle \vec{S}^\psi_j \rangle / |\langle \vec{S}^\psi_j \rangle|$.
At long distances between $\br_j$ and $\br_{j'}$ this function decays exponentially with an exponential dependence $e^{-r/\xi_A}$ of the distance $r$.
Fourier transforming and using the rotated spin frame introduced in Sec.~\ref{sec: spin stiff formalism}, the long distance behavior of $\chi_A(\mathbf{r}_{j},\mathbf{r}_{j'})$ can be related to the momentum dependence of the static correlation function $\chit^{ab}(\bq,0)$ in the amplitude channel $a=b=1$ for small $\bq$, which has the general form
\begin{equation}
 \chit^{11}(\bq,0) \propto \frac{1}{J^A_{\alpha\beta} q_\alpha q_\beta + m_A^2} \, .
\end{equation}
The magnetic coherence length is then given by
\begin{equation}
 \xi_A = \sqrt{J_A}/(2 m_A) \, ,
\end{equation}
where $J_A = \left( J_{xx}^A J_{yy}^A - J_{xy}^A J_{yx}^A \right)^\frac{1}{2}$.

The constant $C$ in Eq.~\eqref{eq: Luv} can be estimated by considering the Hubbard model with pure nearest neighbor hopping (with amplitude $-t$) at half-filling. At strong coupling (large $U$) the spin degrees of freedom are then described by the antiferromagnetic Heisenberg model, which exhibits a N\'eel ordered ground state with a magnetization reduced by a factor $n_0 \approx 0.6$ compared to the mean-field value \cite{Manousakis1991}. On the other hand, evaluating the RPA expressions for the Hubbard model in the strong coupling limit, one recovers the mean-field results for the spin stiffness and spin wave velocity of the Heisenberg model with an exchange coupling $J_H = 4t^2/U$, namely $J = J_H/4$ and $c_s = \sqrt{2} J_H$. Evaluating the RPA spin amplitude correlation function yields $\xi_A = 1/\sqrt{8}$ in this limit. With the ansatz \eqref{eq: Luv}, one then obtains $n_0 = 1 - 4C/\pi$. Matching this with the numerical result $n_0 \approx 0.6$ yields $C \approx 0.3$ and $\Luv \approx 0.9$.

We finally note that we are not overcounting any fluctuations in our theory. In general, the electron fractionalization in Eq.~\eqref{eq: electron fractionaliz.} introduces redundant degrees of freedom associated with the gauge symmetry, Eq.~\eqref{eq: gauge symmetry}. We have not explicitly fixed a gauge but, due to our (renormalized) mean-field treatment of the chargons, fluctuations of the magnetic order parameter are captured exclusively by the spinons.


\section{Results}
In this section we present and discuss results obtained from our theory for the two-dimensional Hubbard model, both in the hole- ($n<1$) and electron-doped ($n>1$) regime. We allow for next and second nearest neighbor hopping with amplitudes $-t$ and $-t'$, respectively, and we fix the ratio of the hopping amplitudes as $t'/t = -0.2$, and we choose a moderate interaction strength $U=4t$. The energy unit is $t$ in all plots. 


\subsection{Chargon mean-field phase diagram}

\begin{figure}[t]
 \centering
 \includegraphics[width=0.48\textwidth]{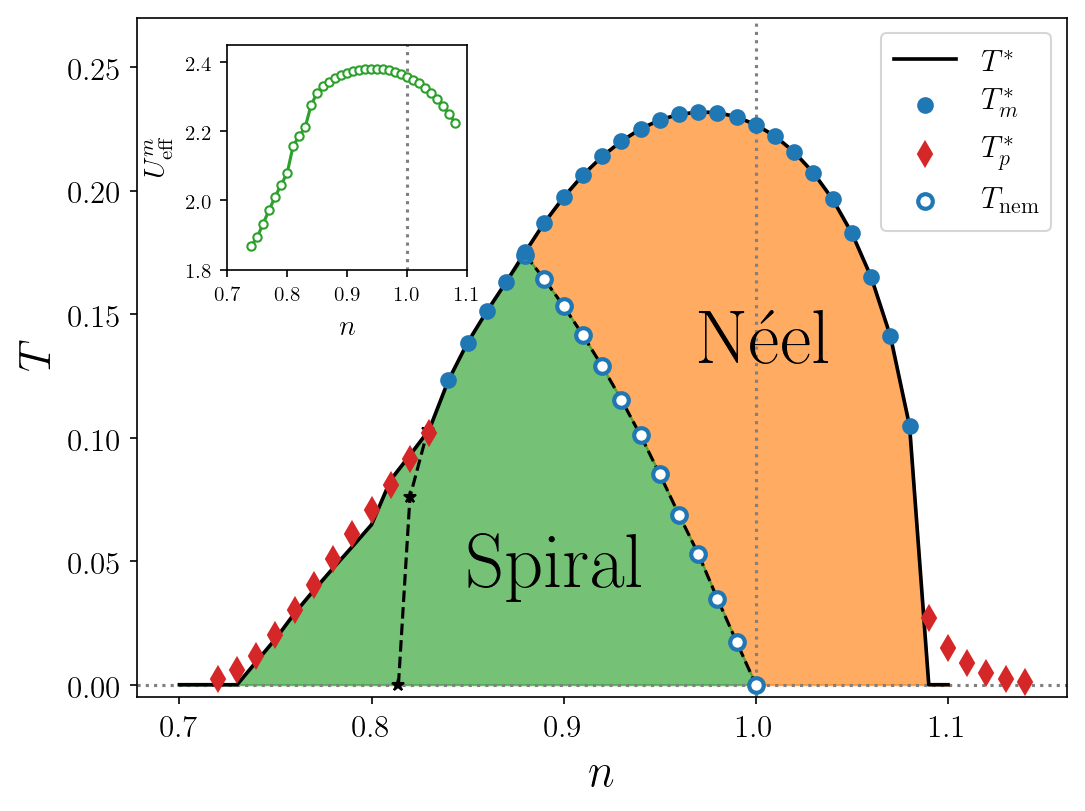}
 \caption{Pseudocritical temperatures $T^*$ and nematic transition temperature $T_{\rm nem}$ as a function of density $n$. The symbols labeled by $T^*_m$ and $T^*_p$ indicate the temperatures at which the effective interaction diverges, in the magnetic or in the pairing channel, respectively. The black solid line labeled by $T^*$ indicates the onset of magnetic order of the chargons and thus the boundary of the pseudogap regime in the absence of superconductivity. $T^*$ coincides with the divergence temperature $T_m^*$ for densities where the vertex diverges in the magnetic channel, and in the hole doped regime it is only slightly lower than $T_p^*$ when the leading divergence occurs in the pairing channel.
 The labels ``N\'eel'' and ``Spiral'' refer to the type of chargon order.
 The dashed black line indicates a topological transition of the quasiparticle Fermi surface within the spiral regime.
 The inset shows the irreducible magnetic effective interaction $U_{\rm eff}^m$ as a function of density.}
\label{fig: Fig2}
\end{figure}
The critical temperatures $T_m^*$ and $T_p^*$ at which the vertex $V^T(\bk_1,\bk_2,\bk_3,\bk_4)$ diverges are shown in Fig.~\ref{fig: Fig2}. In a density range from $n \approx 0.83$ to $n \approx 1.08$, the divergence of the vertex is due to a magnetic instability. Beyond the edges of this density interval, the leading instability occurs in the $d$-wave pairing channel. 
Pairing extends into the magnetic regime at lower temperatures (below $T_m^*$) as a secondary instability. Vice versa, magnetic order is possible at temperatures below $T_p^*$ in the regime where pairing fluctuations dominate \cite{Wang2014, Yamase2016}.

In Fig.~\ref{fig: Fig2}, we also show the irreducible effective magnetic interaction $U_\mathrm{eff}^m$ defined in Sec.~\ref{sec: order parameter and Q}. The effective interaction $U_\mathrm{eff}^m$ is strongly reduced from its bare value ($U=4t$) by the non-magnetic channels in the fRG flow, while its density dependence is not very strong.

\begin{figure}[t]
 \centering
 \includegraphics[width=0.48\textwidth]{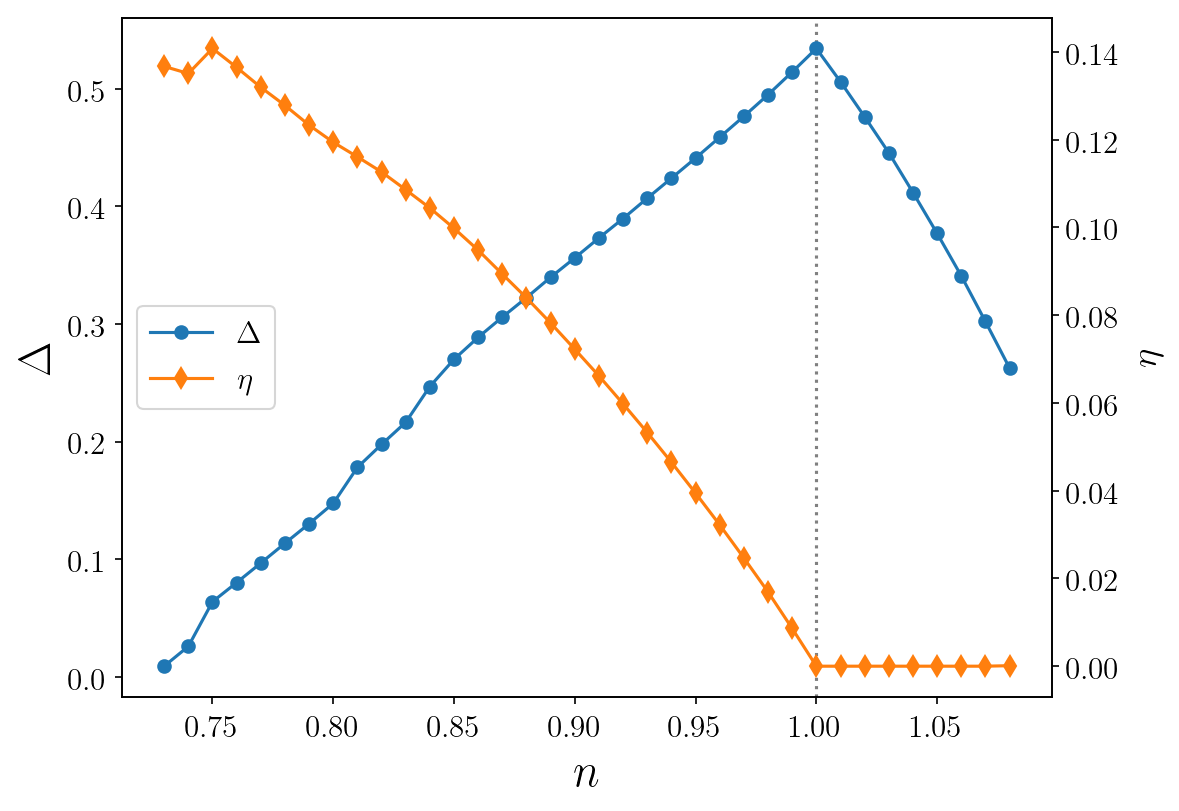}
 \caption{Magnetic gap $\Delta$ (left axis) and incommensurability $\eta$ (right axis) at $T=0$ as functions of the density.}
 \label{fig: Fig3}
\end{figure}
From now on we ignore the pairing instability and focus on the magnetic order of the chargons. We compute the magnetic order parameter $\Delta$ together with the optimal wave vector $\bQ$ as described in Sec.~\ref{sec: order parameter and Q}. In Fig.~\ref{fig: Fig3}, we show results for $\Delta$ in the ground state ($T=0$) as a function of the filling. We find a stable magnetic solution extending deep into the hole doped regime down to $n \approx 0.73$. On the electron doped side magnetic order terminates abruptly already at $n \approx 1.08$. This pronounced electron-hole asymmetry and the discontinuous transition on the electron doped side has already been observed in previous fRG+MF calculations for a slightly weaker interaction $U = 3t$ \cite{Yamase2016}.

The onset temperature $T^*$ for magnetic order of the chargons as obtained from the renormalized mean-field theory is shown in Fig.~\ref{fig: Fig2}.
At densities where magnetic interactions dominate, it coincides with the temperature $T_m^*$ at which $V^T$ diverges. At densities where the interaction diverges in the pairing channel, $T^*$ is lying only slightly below $T_p^*$ on the hole doped side, while it vanishes on the electron doped side.
While the magnetic gap in the ground state reaches its peak at $n=1$, as expected, the pseudocritical temperature $T^*$ and the irreducible effective interaction $U_{\rm eff}^m$ exhibit their maximum in the hole doped regime slightly away from half-filling.

The magnetic states are either N\'eel type or spiral with a wave vector of the form $\bQ = (\pi-2\pi\eta,\pi)$, or symmetry related, with an ``incommensurability'' $\eta > 0$. 
In Fig.~\ref{fig: Fig3} results for $\eta$ in the ground state are shown as a function of the density. At half-filling and in the electron doped region only N\'eel order is found, as expected and in agreement with previous fRG+MF studies \cite{Yamase2016}. Hole doping instead immediately leads to a spiral ground state with $\eta > 0$. Whether the N\'eel state persists at small hole doping depends on the hopping parameters and the interaction strength. Its instability toward a spiral state is favored by a larger interaction strength \cite{Chubukov1995}. Indeed, in a previous fRG+MF calculation at weaker coupling the N\'eel state was found to survive up to about 10 percent hole doping \cite{Yamase2016}.

At low and moderate hole doping, there is a transition between a N\'eel state at high temperatures and a spiral state at low temperatures. Since the spiral state breaks the tetragonal symmetry of the square lattice, spiral order entails electronic nematicity. In Fig.~\ref{fig: Fig2} we show the corresponding nematic transition temperature $T_{\rm nem}$ as a function of density. $T_{\rm nem}$ merges with $T^*$ at $n \approx 0.88$. For lower densities the magnetic order is spiral with $\eta > 0$ at any temperature below the magnetic transition temperature.
Within the spiral regime there is a topological transition of the quasiparticle Fermi surface (indicated by the black dashed line in Fig.~\ref{fig: Fig2}), where hole pockets merge. The Fermi surface extracted from the single-particle spectral function develops Fermi arcs on the right hand side of this transition, while it resembles the large bare Fermi surface on the left (see Sec.~\ref{sec: specfct}).


\subsection{Spinon fluctuations}

\begin{figure*}[t]
\centering
 \includegraphics[width=0.87\textwidth]{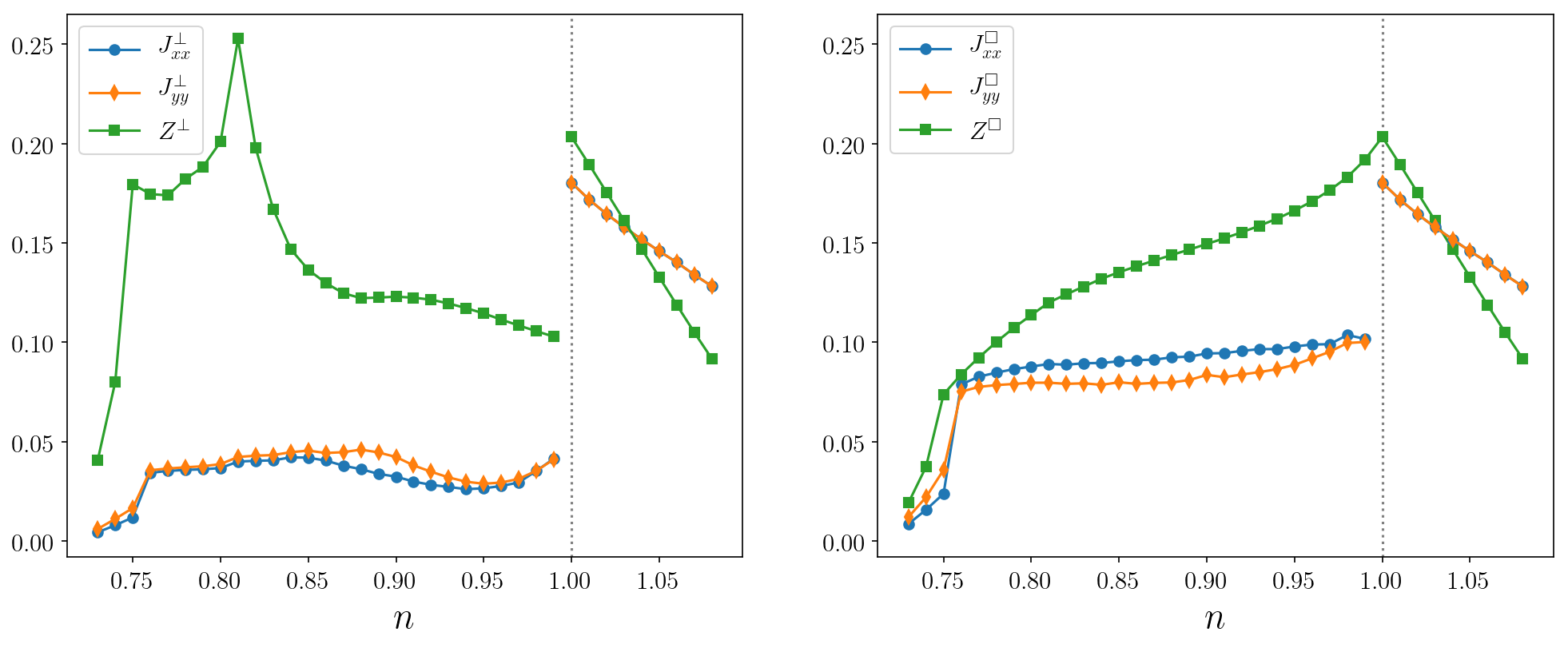}
 \caption{Out-of-plane (left panel) and in-plane (right panel) spatial ($J$) and temporal ($Z$) spin stiffnesses in the ground state ($T=0$) as functions of the filling $n$. In the N\'eel state (for $n \geq 1$) out-of-plane and in-plane stiffnesses coincide.}
\label{fig: Fig4}
\end{figure*}
Once the magnetic order parameter $\Delta$ of the chargons and the wave vector $\bQ$ have been computed, we are in the position to calculate the NL$\sigma$M parameters from the expressions presented in Sec.~\ref{sec: spin stiff formalism}. 

In Fig.~\ref{fig: Fig4}, we plot results for the spatial and temporal spin stiffnesses $J^a_{\alpha\alpha}$ and $Z^a$ in the ground state. In the spiral state (for $n < 1$) out-of-plane and in-plane stiffnesses are distinct, while in the N\'eel state (for $n \geq 1$) they coincide. Actually the order parameter defines an axis, not a plane, in the latter case.
All the quantities except $Z^\Box$ exhibit pronounced jumps between half-filling and infinitesimal hole-doping. These discontinuities are due to the appearance of hole pockets around the points $(\frac{\pi}{2},\frac{\pi}{2})$ in the Brillouin zone \cite{Bonetti2022}. The spatial stiffnesses are almost constant over a broad range of hole-doping, with a small spatial anisotropy $J^a_{xx} \neq J^a_{yy}$. The temporal stiffnesses $Z^a$ exhibit a stronger doping dependence. The peak of $Z^\perp$ at $n \approx 0.79$ is associated with a van Hove singularity of the quasiparticle dispersion \cite{Bonetti2022}. On the electron doped side all stiffnesses decrease almost linearly with the electron filling. The off-diagonal spin stiffnesses $J^a_{xy}$ and $J^a_{yx}$ vanish both in the N\'eel state and in the spiral state with $\bQ = (\pi-2\pi\eta,\pi)$ and symmetry related.

\begin{figure}[t]
\centering
 \includegraphics[width=0.48\textwidth]{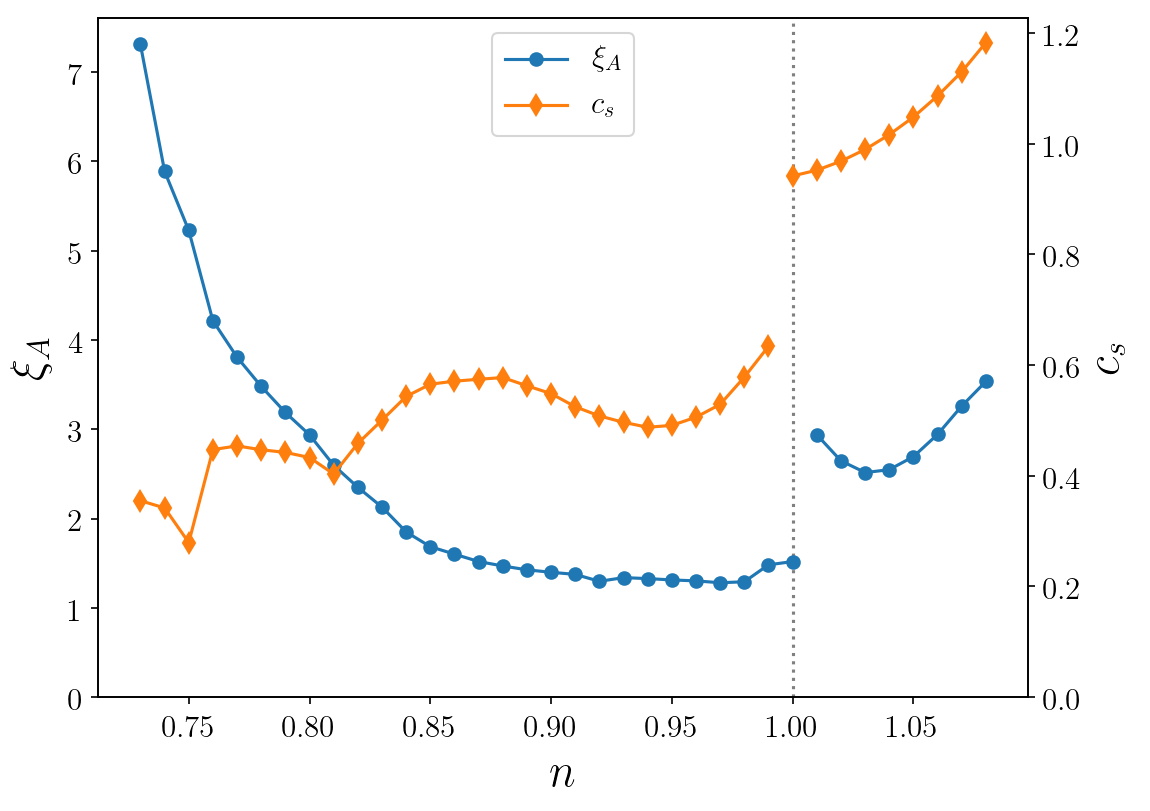}
 \caption{Magnetic coherence length $\xi_A$ (left axis) and average spin wave velocity $c_s$ in the ground state as functions of the filling $n$.}
\label{fig: Fig5}
\end{figure}
In Fig.~\ref{fig: Fig5} we show the magnetic coherence length $\xi_A$ and the average spin wave velocity $c_s$ in the ground state. The coherence length is rather short and only weakly doping dependent from half-filling up to 15 percent hole-doping, while it increases strongly toward the spiral-to-paramagnet transition on the hole-doped side. On the electron-doped side it almost doubles from half-filling to infinitesimal electron doping. This jump is due to the formation of electron pockets upon electron doping. Note that $\xi_A$ does not diverge at the transition to the paramagnetic state on the electron doped side, as this transition is first order.
The average spin wave velocity exhibits a pronounced jump at half-filling, which is inherited from the jumps of $J_{\alpha\alpha}^\perp$ and $Z^\perp$. Besides this discontinuity it does not vary much as a function of density.

We now investigate whether the magnetic order in the ground state is destroyed by quantum fluctuations or not. To this end we compute the boson condensation fraction $n_0$ as obtained from the large-$N$ expansion of the NL$\sigma$M. This quantity depends on the ultraviolet cutoff $\Luv$. As a reference point, we may use the half-filled Hubbard model at strong coupling, as discussed in Sec.~\ref{sec: cutoff}, which yields $\Luv \approx 0.9$, and the constant in the ansatz Eq.~\eqref{eq: Luv} is thereby fixed to $C \approx 0.3$.

\begin{figure}[t]
\centering
\includegraphics[width=0.48\textwidth]{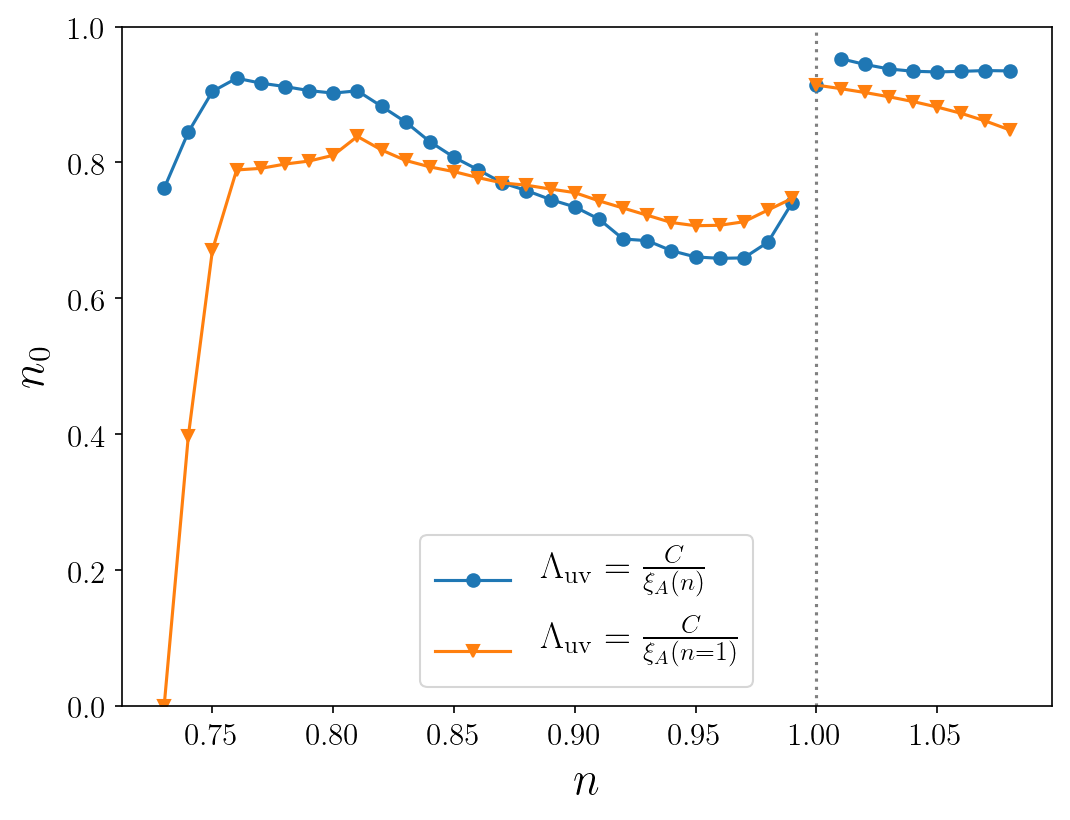}
 \caption{Fraction of condensed $z$-bosons $n_0$ at $T=0$ for two distinct choices of the ultraviolet cutoff $\Luv$ as a function of the filling.}
\label{fig: Fig6}
\end{figure}
In Fig.~\ref{fig: Fig6} we show the condensate fraction $n_0$ computed with two distinct choices of the ultraviolet cutoff: $\Luv = \Luv(n) = C/\xi_A(n)$ and $\Luv = C/\xi_A(n=1)$. For the former choice the cutoff vanishes at the edge of the magnetic region on the hole-doped side, where $\xi_A$ diverges. One can see that $n_0$ remains finite for both choices of the cutoff in nearly the entire density range where the chargons order. Only near the hole-doped edge of the magnetic regime, $n_0$ vanishes slightly above the mean-field transition point, if the ultraviolet cutoff is chosen as density independent. The discontinuous drop of $n_0$ upon infinitesimal hole doping is due to the corresponding drop of the out-of-plane stiffness, while the discontinuous increase of $n_0$ upon infinitesimal electron doping, for the density dependent cutoff choice $\Luv(n) = C/\xi_A(n)$, is due to the discontinuity of $\xi_A(n)$. In the weakly hole-doped region there is a substantial reduction of $n_0$ below one, for both choices of the cutoff. Except for the edge of the magnetic region on the hole-doped side, the choice of the cutoff has only a mild influence on the results, and the condensate fraction remains well above zero. Hence, we can conclude that the ground state of the Hubbard model with a moderate coupling $U = 4t$ is magnetically ordered over wide density range. The spin stiffness is sufficiently large to protect the magnetic order against quantum fluctuations of the order parameter. 


\subsection{Electron spectral function}
\label{sec: specfct}

Fractionalizing the electron operators as in Eq.~\eqref{eq: electron fractionaliz.}, the electron Green's function assumes the form
\begin{eqnarray}
 [\cG^e_{jj'}(\tau)]_{\sigma\sigma'} &=&
 - \langle c_{j'\sigma'}(\tau) c^*_{j\sigma}(0) \rangle \nonumber \\
 &=& - \langle [R_{j'}(\tau)]_{\sigma' s'}[R_j^*(0)]_{\sigma s} \,
 \psi_{j's'}(\tau) \psi^*_{js}(0) \rangle \, . \nonumber \\
\end{eqnarray}
To simplify this expression, we decouple the average $\langle R R^* \psi \psi^* \rangle$ as $\langle R R^* \rangle \langle \psi \psi^* \rangle$, yielding \cite{Borejsza2004, Scheurer2018, Wu2018}
\begin{equation} \label{eq: Ge decoupled}
 [\cG^e_{jj'}(\tau)]_{\sigma\sigma'} =
 - \langle[R_{j'}(\tau)]_{\sigma' s'}[R_j^*(0)]_{\sigma s} \rangle
 \langle \psi_{j's'}(\tau) \psi_{js}^*(0) \rangle \, .
\end{equation}
The spinon Green's function can be computed from the NL$\sigma$M in the continuum limit. Using the Schwinger boson parametrization~\eqref{eq: R to z}, we obtain, in the large $N$ limit,
\begin{eqnarray}
 \langle [R(\br_{j'},\tau)]_{\sigma' s'} [R^*(\br_j,0)]_{\sigma s} \rangle &=&
 - D(\br_j \!-\! \br_{j'},\tau) \, \delta_{\sigma\sigma'} \delta_{ss'}  \nonumber \\
 && + \, n_0 \, \delta_{\sigma s} \delta_{\sigma' s'} \, .
\end{eqnarray}
The boson propagator $D(\br,\tau)$ is the Fourier transform of
\begin{equation}
 D(\bq,\omega_n) =
 \frac{1}{Z^\perp \omega_n^2 + J_{\alpha\beta}^\perp q_\alpha q_\beta + Z^\perp m_s^2} \, ,
\end{equation}
with the bosonic Matsubara frequency $\omega_n = 2\pi n T$.
Fourier transforming Eq.~\eqref{eq: Ge decoupled}, the electron Green's function is obtained in momentum representation as
\begin{eqnarray} \label{eq: Ge nu_n}
 \cG^e(\bk,\bk',\nu_n) &=& - T \sum_{\omega_m} \int_\bq 
 \tr \left[ \cG(\bk-\bq,\bk'-\bq,\nu_n - \omega_m) \right] \nonumber \\
 && \times D(\bq,\omega_m) \, \mathbb{1} + n_0 \, \cG(\bk,\bk',\nu_n) \, ,
\end{eqnarray}
where $\cG(\bk,\bk',\nu_n)$ is the chargon Green's function.

We see that when $n_0 = 0$, the electron Green's function is diagonal in momentum, that is, it is translational invariant, as the diagonal components of the chargon Green's function entering the trace are nonzero only for $\bk=\bk'$. Furthermore, in this case $\cG^e$ is proportional to the unity matrix in spin space, since there is no spin SU(2) symmetry breaking, and is thus given by a single normal state Green's function $G^e(\bk,\nu_n)$.
Performing the Matsubara sum in Eq.~\eqref{eq: Ge nu_n} and continuing to real frequencies, we get
\begin{eqnarray}
 G^e(\bk,\omega) &=& \sum_{\ell=\pm} \sum_{p=\pm} \int_{|\bq|\leq\Luv} 
 \frac{1}{4Z^\perp \omega^{\mathrm{sp}}_\bq} \left(1+\ell\frac{h_{\bk-\bq}}{e_{\bk-\bq}}
 \right) \nonumber \\
 &\times& \frac{f(pE^\ell_{\bk-\bq}) + n_B(\omega^{\mathrm{sp}}_\bq)}
 {\omega+i0^+-E^\ell_{\bk-\bq}+p\,\omega^{\mathrm{sp}}_\bq} + \{\bk\to-\bk\} \, , \hskip 7mm
\end{eqnarray}
where 
\begin{equation}
 \omega^\mathrm{sp}_\bq = \sqrt{(J^\perp_{\alpha\beta} q_\alpha q_\beta)/Z^\perp + m_s^2} \, ,
\end{equation}
and $n_B(x)=(e^{x/T}-1)^{-1}$ is the Bose distribution function.

\begin{figure}[t]
\centering
\includegraphics[width=0.48\textwidth]{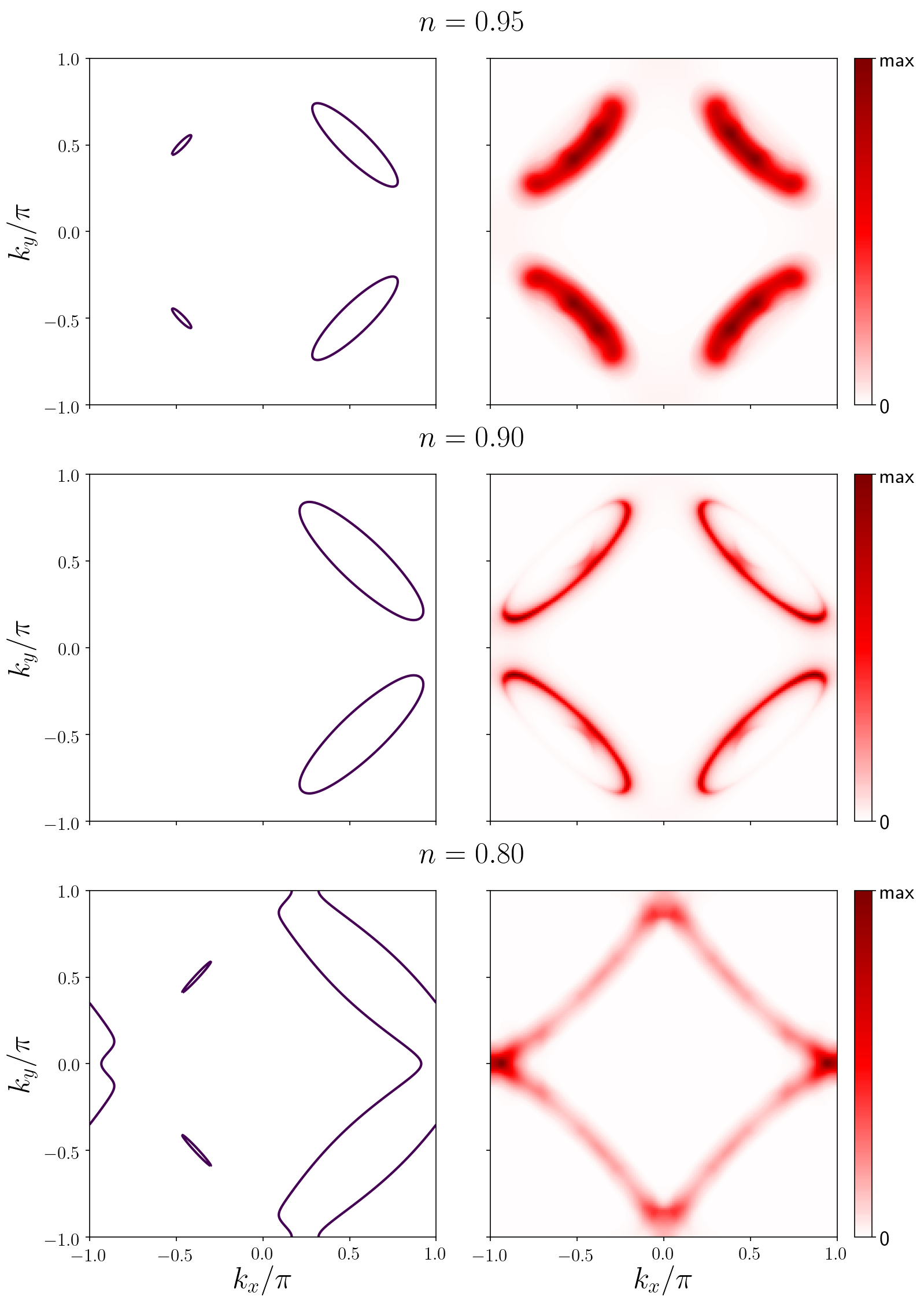}
 \caption{Quasiparticle Fermi surfaces defined as zeros of the chargon quasiparticle energies $E_\bk^\pm$ (left column) and momentum dependence of electron spectral function at zero frequency (right column) for various electron densities. The temperature is $T = 0.05t$.}
\label{fig: Fig7}
\end{figure}
In the right column of Fig.~\ref{fig: Fig7} we show the spectral function obtained as the imaginary part of the retarded electron Green's function at zero frequency as a function of momentum for various electron densities in the hole doped regime. The temperature $T = 0.05t$ is below the chargon ordering temperature in all cases. The Fermi surface topology is the same as the one obtained from a mean field approximation of spiral spin density wave order \cite{Eberlein2016}. At low hole doping it originates from a superposition of hole pockets (see left column of Fig.~\ref{fig: Fig7}), where the spectral weight on the back sides is drastically suppressed by coherence factors, so that only the front sides are visible. The spinon fluctuations lead to a broadening of the spectral function, so that the Fermi surface is smeared out. Since the spinon propagator does not depend on the fermionic momentum, the broadening occurs uniformly in the entire Brillouin zone. Hence, the backbending at the edges of the ``arcs'' obtained in our theory for $n=0.9$ is more pronounced than experimentally observed in cuprates. The backbending edges could be further suppressed by including a momentum dependent self-energy which has a larger imaginary part in the antinodal region \cite{Mitscherling2021}.


\section{Conclusions}

We have presented a SU(2) gauge theory of fluctuating magnetic order in the two-dimensional Hubbard model. The theory is based on a fractionalization of the electron operators in chargons and spinons \cite{Schulz1995, Dupuis2002, Borejsza2004, Sachdev2009}. The chargons are treated in a renormalized mean-field theory with effective interactions obtained from a functional renormalization group flow. They undergo N\'eel or spiral magnetic order in a broad density range around half-filling below a density dependent temperature $T^*$. Fluctuations of the spin orientation are described by a non-linear sigma model obtained from a gradient expansion of the spinon degrees of freedom. The parameters of the sigma model, the spin stiffnesses, have been computed from a renormalized RPA. Our approximations are applicable for a weak or moderate Hubbard interaction $U$. While magnetic long-range order of the electrons is still possible in the ground state, at any finite temperature the spinon fluctuations prevent long-range order -- in agreement with the Mermin-Wagner theorem. We expect that at strong coupling even the ground state becomes disordered already at relatively low hole-doping, since fluctuations are then enhanced due to the shorter magnetic coherence length.

In spite of the moderate interaction strength chosen in our explicit calculations, the phase with magnetic chargon order below $T^*$ exhibits all important features characterizing the pseudogap regime in high-$T_c$ cuprates. The Fermi surface reconstruction yields a reduction of the electronic density of states. At low hole doping the Fermi surface obtained from the spectral function for single-particle excitations looks like Fermi arcs. The spinon fluctuations generate a spin gap at any finite temperature. The spinon fluctuations do not contribute to quantities involving only charge degrees of freedom, such as the longitudinal or Hall conductivities. It was already shown previously that N\'eel or spiral order of the chargons can explain the drastic charge carrier drop observed at the onset of the pseudogap regime in hole-doped cuprates \cite{Storey2016, Storey2017, Eberlein2016, Chatterjee2017a, Verret2017, Mitscherling2018, Bonetti2020_I}.

In the N\'eel regime, the structure of our theory is very similar to the SU(2) gauge theory of the pseudogap phase derived by Sachdev and coworkers \cite{Sachdev2009, Sachdev2016, Chatterjee2017, Scheurer2018, Wu2018, Sachdev2019}. Besides our extension to spiral states, the major new aspect of our work is that we \emph{compute} the magnetic order parameter and spin stiffnesses instead of fitting the parameters of the theory. This computation revealed in particular an important particle-hole asymmetry of the stiffnesses.

Spiral order of the chargons entails nematic order of the electrons. At low hole doping, the chargons form a N\'eel state at $T^*$, and a spiral state at a lower temperature $T_{\rm nem}$. The electrons thus undergo a nematic phase transition at a critical temperature {\em below}\/ the pseudogap temperature. Evidence for a nematic transition at a temperature $T_{\rm nem} < T^*$ has been found recently in slightly underdoped YBCO \cite{Grissonnanche2022}. For large hole doping instead, the nematic transition occurs right at $T^*$, while nematic order is completely absent for electron doping, that is, above half-filling.

In the ground state of the two-dimensional Hubbard model there is a whole zoo of possible magnetic ordering patterns, and away from half filling N\'eel or spiral order do not always minimize the ground state energy. The most important competitor is stripe order, that is, collinear spin order associated with charge order, where holes accumulate in one-dimensional lines \cite{Qin2022}. Stripe order in the ground state has been established rather convincingly for special cases, such as pure nearest neighbor hopping and doping concentration $1/8$ \cite{Zheng2017}. The energy difference between distinct order patterns can be very small.
At finite temperatures, the issue of the proper choice of the magnetic order reappears for the chargons. A classification of the numerous possibilities has been provided recently by Sachdev et al. \cite{Sachdev2019a}. We have focused on N\'eel and spiral states because any other state leads to a fractionalization of the Fermi surface into numerous tiny pieces (infinitly many for incommensurate wave vectors), which is in conflict with the experimental observation of only four arcs in the pseudogap phase of cuprates. Moreover, it is hard to explain the sharp carrier drop observed at the edge of pseudogap regime in high magnetic fields via collinear magnetic order \cite{Charlebois2017}. Hence, to us N\'eel or spiral order of the chargons seems the most promising starting point to understand the universal features of the pseudogap phase. Refinements are required to capture also secondary instabilities, that is, charge order and superconductivity.

At finite temperatures, we obtain a ``pseudogap'' phase with a reconstructed Fermi surface and a spin gap also for the electron doped Hubbard model. In contrast, in electron doped cuprates one observes a comparatively broad (in doping) N\'eel phase, and no or only a very narrow pseudogap regime. N\'eel order at finite temperature is possible due to the interlayer coupling in cuprates. On the hole doped side, interlayer coupling stabilizes the N\'eel state only in a very narrow regime near half-filling. This electron-hole asymmetry can be explained by the asymmetry of the spin stiffnesses, which are much smaller on the hole doped side (see Fig.~\ref{fig: Fig4}), enhancing thus the impact of spin fluctuations.


\section*{Acknowledgements}

We are very grateful to Andres Greco, Elio K\"onig and Demetrio Vilardi for valuable discussions.


\appendix

\section{Linear term in the gauge field}
\label{app: linear term}
In this Appendix we show that the linear term in Eq.~\eqref{eq: effective action Amu} vanishes. Fourier transforming the vertex and the expectation value, the coefficient $\mathcal{B}_\mu^a$ can be written as
\begin{equation}
 \mathcal{B}_\mu^a = \frac{1}{2} \int_\bk T \sum_{\nu_n} \gamma_\mu^{(1)}(\bk)
 \Tr \left[ \sigma^a \cG(\bk,\bk,\nu_n) \right] \, .
\end{equation}
Inserting $\cG$ from Eq.~\eqref{eq: calG} one immediately sees that $\mathcal{B}_\mu^1 = \mathcal{B}_\mu^2 = 0$ for $\mu = 0,1,2$, and $\mathcal{B}_0^3 = 0$, too. 
Performing the Matsubara sum for $\mathcal{B}^3_\alpha$ with $\alpha = 1,2$, we obtain
\begin{equation}
 \mathcal{B}^3_\alpha = \frac{1}{2} \int_\bk
 \sum_{\ell=\pm} \left[
 (\partial_{k_\alpha}\epsilon_\bk) u^\ell_\bk f(E^\ell_\bk) + 
 (\partial_{k_\alpha}\epsilon_{\bk+\bQ}) u^{-\ell}_\bk f(E^\ell_\bk) \right] \, ,
\label{eq: linear term in Amu expression}
\end{equation}
where $u^\ell_\bk = \frac{1}{2} \big( 1 + \ell h_\bk/\sqrt{h_\bk^2 + \Delta^2} \big)$
with $h_\bk = \frac{1}{2} (\epsilon_\bk - \epsilon_{\bk+\bQ})$.
One can see by direct calculation that this term vanishes if $\partial F(\bQ)/\partial\bQ$ with $F(\bQ)$ given by Eq.~\eqref{eq: MF theromdynamic potential} vanishes. Hence, $\mathcal{B}^3_\alpha$ vanishes if $\bQ$ minimizes the free energy.
A similar result has been obtained in Ref.~\cite{Klee1996}. 
%


\section{Derivation of the NL\texorpdfstring{$\sigma$}{s}M}
\label{app: derivation of the NLsM}
Here we derive the NL$\sigma$M action~\eqref{eq: general NLsM} from Eq.~\eqref{eq: effective action Amu}. We first prove the identity
\begin{equation} \label{eq: dmu R identity}
 \dmu\cR = -i \cR\, \Sigma^a A_\mu^a,
\end{equation}
where $\cR$ is defined by Eq.~\eqref{eq: R to mathcal R}, and $\Sigma^a$ are the generators of the SU(2) in the adjoint representation,
\begin{equation}
 \Sigma^a_{bc} = -i \varepsilon^{abc},
\end{equation}
with $\varepsilon^{abc}$ the Levi-Civita tensor. 
Rewriting Eq.~\eqref{eq: R to mathcal R} as 
\begin{equation}
 \cR^{ab} = \frac{1}{2} 
 \Tr\left[ R^\dagger \sigma^a R^{\phantom{\dagger}} \sigma^b \right] \, ,
\end{equation}
we obtain the derivative of $\cR$ in the form,
\begin{eqnarray}
 \dmu\cR^{ab} &=& \Tr\left[ R^\dagger \sigma^a \, (\dmu R) \sigma^b \right] =
 \Tr\left[R^\dagger \sigma^a R R^\dagger(\dmu R) \sigma^b \right] \nonumber \\
 &=& -i \cR^{ac} \Sigma^d_{cb} A_\mu^d,
\end{eqnarray}
which is the identity in \eqref{eq: dmu R identity}.

We now aim to express the object $\frac{1}{2} \cJ^{ab}_\munu A_\mu^\a A^b_\nu$ in terms of the matrix field $\cR$. We write the stiffness matrix in terms of a new matrix $\cP_\munu$ via
\begin{equation} \label{eq: J to P}
 \cJ^{ab}_\munu =\Tr[ \cP_\munu ]\delta_{ab} - \cP^{ab}_\munu  = 
 \Tr \left[ \cP_\munu \Sigma^a \Sigma^b \right] .
\end{equation}
Using $\cR^T \cR = \mathbb{1}$, we obtain
\begin{eqnarray}
 \frac{1}{2} \cJ^{ab}_\munu A_\mu^a A^b_\nu &=& 
 \frac{1}{2} \Tr \left[ \cP_\munu \,
 \Sigma^a \, \cR^T \cR \, \Sigma^b \right] A_\mu^a A^b_\nu \nonumber \\
 &=& \frac{1}{2} \Tr \left[ \cP_\munu (\dmu\cR^T)(\dnu\cR) \right] ,
\end{eqnarray}
where we have used Eq.~\eqref{eq: dmu R identity} in the last line. The above equation yields
Eq.~\eqref{eq: general NLsM}. Relation~\eqref{eq: J to P} can be easily inverted using $\Tr[\cJ_\munu ] = 2\Tr[\cP_\munu]$.


\section{Details on the large-\texorpdfstring{$N$}{N} expansion}
\label{app: large N}

In this Appendix, we describe some details regarding the saddle point equations of the CP$^{N-1}$ action. Integrating out the $z$-bosons from Eq.~\eqref{eq: massive CPN1 model}, we obtain the effective action \cite{AuerbachBook1994}
\begin{eqnarray}
 \cS[\cA_\mu,\lambda] &=& N \int_\mathcal{T} dx \Big[
 \ln \left( -2J^\perp_{\mu\nu} D_\mu D_\nu + i \lambda \right)
 - \frac{i}{2} \lambda \nonumber \\
 && + \,\frac{1}{4} M_{\mu\nu} \cA_\mu \cA_\nu \Big] \, . 
\end{eqnarray}
In the large $N$ limit the functional integral for its partition function
is dominated by its saddle point, which is determined by the stationarity equations
\begin{equation}
 \frac{\delta\mathcal{S}}{\delta\mathcal{A}_\mu}= 
 \frac{\delta\mathcal{S}}{\delta\lambda} = 0 \, .
\end{equation}

The first condition implies $\mathcal{A}_\mu=0$, that is, in the large-$N$ limit the U(1) gauge field fluctuations are totally suppressed. The variation with respect to $\lambda$ gives, assuming a spatially uniform average value for $\lambda$, 
\begin{equation}
 n_0 + T \sum_{\omega_n} \int_\bq
 \frac{1}{Z^\perp \omega_n^2 + J^\perp_{\alpha\beta} q_\alpha q_\beta +
 i\langle\lambda\rangle} = 1 \, ,
\end{equation}
where $n_0$ is the fraction of condensed bosons, which can be nonzero at $T=0$.
Performing the sum over the bosonic Matsubara frequencies $\omega_n = 2n\pi T$, inserting the identity 
\begin{equation}
 1 = \int_0^\infty \! d\epsilon \,
 \delta\Big( \epsilon - \sqrt{ J^\perp_{\alpha\beta} q_\alpha q_\beta/Z^\perp } \, \Big) ,
\end{equation}
and performing the $\bq$-integral, we obtain Eq.~\eqref{eq: large N equation} at $T > 0$ and Eq.~\eqref{eq: large N eq at T=0} at $T=0$.


\bibliography{biblio.bib}

\end{document}